%% file: schubert.tex
\documentclass[11pt]{article}
\usepackage{graphicx,eepic,epic,epsfig,amsmath,latexsym,amssymb,amscd}
\usepackage{youngtab,verbatim}
\usepackage[square,numbers,sort&compress]{natbib}

\input xy
\xyoption{all}

\usepackage{theorem}
\newtheorem{definition}{Definition}[section]
\newtheorem{proposition}[definition]{Proposition}
\newtheorem{lemma}[definition]{Lemma}

\newtheorem{theorem}[definition]{Theorem}
\newtheorem{corollary}[definition]{Corollary}
\newtheorem{conjecture}[definition]{Conjecture}

\newtheorem{observation}[definition]{Observation}

\def\squareforqed{\hbox{\rlap{$\sqcap$}$\sqcup$}}
\def\qed{\ifmmode\squareforqed\else{\unskip\nobreak\hfil
\penalty50\hskip1em\null\nobreak\hfil\squareforqed
\parfillskip=0pt\finalhyphendemerits=0\endgraf}\fi}
\def\endenv{\ifmmode\;\else{\unskip\nobreak\hfil
\penalty50\hskip1em\null\nobreak\hfil\;
\parfillskip=0pt\finalhyphendemerits=0\endgraf}\fi}

\mathchardef\ordinarycolon\mathcode`\:
\mathcode`\:=\string"8000
\def\vcentcolon{\mathrel{\mathop\ordinarycolon}}
\begingroup \catcode`\:=\active
  \lowercase{\endgroup
  \let :\vcentcolon
  }

\newcommand{\captionfonts}{\small}
\makeatletter  
\long\def\@makecaption#1#2{%
  \vskip\abovecaptionskip
  \sbox\@tempboxa{{\captionfonts \textbf{#1}\; #2}}%
  \ifdim \wd\@tempboxa >\hsize
    {\captionfonts \textbf{#1}\; #2\par}
  \else
    \hbox to\hsize{\hfil\box\@tempboxa\hfil}%
  \fi
  \vskip\belowcaptionskip}
\makeatother   

\newcommand{\nc}{\newcommand}
\nc{\rnc}{\renewcommand}
\nc{\beq}{\begin{equation}}
\nc{\eeq}{{\end{equation}}}
\nc{\beqa}{\begin{eqnarray}}
\nc{\eeqa}{\end{eqnarray}}
\nc{\lbar}[1]{\overline{#1}}
\nc{\bra}[1]{\langle#1|}
\nc{\ket}[1]{|#1\rangle}
\nc{\ketbra}[2]{|#1\rangle\!\langle#2|}
\nc{\braket}[2]{\langle#1|#2\rangle}
\nc{\proj}[1]{| #1\rangle\!\langle #1 |}
\nc{\avg}[1]{\langle#1\rangle}
\rnc{\max}{\operatorname{max}}
\nc{\Rank}{\operatorname{Rank}}
\nc{\smfrac}[2]{\mbox{$\frac{#1}{#2}$}}
\nc{\Tr}{\operatorname{Tr}}
\nc{\ox}{\otimes}
\nc{\dg}{\dagger}
\nc{\dn}{\downarrow}
\nc{\cA}{{\cal A}}
\nc{\cB}{{\cal B}}
\nc{\cC}{{\cal C}}
\nc{\cD}{{\cal D}}
\nc{\cE}{{\cal E}}
\nc{\cF}{{\cal F}}
\nc{\cG}{{\cal G}}
\nc{\cH}{{\cal H}}
\nc{\cI}{{\cal I}}
\nc{\cJ}{{\cal J}}
\nc{\cK}{{\cal K}}
\nc{\cO}{{\cal O}}
\nc{\cL}{{\cal L}}
\nc{\cP}{{\cal P}}
\nc{\cR}{{\cal R}}
\nc{\cS}{{\cal S}}
\nc{\cT}{{\cal T}}
\nc{\cX}{{\cal X}}
\nc{\cZ}{{\cal Z}}
\nc{\rar}{\rightarrow}
\nc{\lrar}{\longrightarrow}
\nc{\Spec}{\operatorname{Spec}}
\nc{\mfk}{\mathfrak k}
\nc{\mfkd}{\mathfrak k^*}
\nc{\GL}{\mbox{GL}}
\nc{\Ad}{\mbox{Ad}}
\nc{\PGL}{\mbox{PGL}}

\def\a{\alpha}
\def\b{\beta}
\def\g{\gamma}
\def\d{\delta}

\def\h{\eta}

\def\l{\lambda}
\def\m{\mu}
\def\n{\nu}
\def\x{\xi}
\def\p{\pi}
\def\r{\rho}
\def\s{\sigma}

\def\ph{\varphi}
\def\c{\chi}

\def\o{\omega}

\def\L{\Lambda}

\def\Ph{\Phi}

\def\O{\Omega}

\def\Gr{\mbox{Gr}}
\nc\Fl{\mbox{Fl}}

\nc{\RR}{{{\mathbb R}}}
\nc{\CC}{{{\mathbb C}}}
\nc{\FF}{{{\mathbb F}}}
\nc{\NN}{{{\mathbb N}}}
\nc{\ZZ}{{{\mathbb Z}}}
\nc{\PP}{{{\mathbb P}}}
\nc{\QQ}{{{\mathbb Q}}}
\nc{\UU}{{{\mathbb U}}}
\nc{\EE}{{{\mathbb E}}}

\begin{document}

\title{{\Large\bf Quantum State
Transformations and the Schubert Calculus}}

\author{
 Sumit Daftuar$^*$ and Patrick Hayden$^{*,\dg}$ \\
 {\small $^*$Institute for Quantum Information, Caltech 107--81, Pasadena, CA 91125,
 USA}\\
 {\small $^\dg$ School of Computer Science, McGill University, Montreal, QC H3A 2A7
 Canada}
}

\date{16 September 2004}

\maketitle

\begin{center}
{\bf Abstract}
\end{center}
Recent developments in mathematics have provided powerful tools
for comparing the eigenvalues of matrices related to each other
via a moment map. In this paper we survey some of the more
concrete aspects of the approach with a particular focus on
applications to quantum information theory. After discussing the
connection between Horn's Problem and Nielsen's Theorem, we move
on to characterizing the eigenvalues of the partial trace of a
matrix.

\section{Introduction}

This paper presents some applications of recently minted matrix
eigenvalue inequalities to problems in quantum information theory.
Much of the machinery necessary to understand the results is
developed on the fly, with the intention that physicists
unfamiliar with the relevant algebraic topology, representation
theory and symplectic geometry can use this paper as an
introduction to the mathematics. Likewise, while much of the
mathematical material will be familiar to experts, we hope that
they will enjoy seeing their techniques put to use in the service
of a relatively novel application, analysis of the nonlocal
structure of quantum states.

One way of understanding quantum information theory is as the
identification of the basic resources useful for manipulating
information in a quantum-mechanical world, and the pursuit of
optimal methods for converting between these resources. (See
\cite{BS04} for a summary of recent work in the field.) In this
framework, communication of classical and quantum bits (qubits)
are resources, as are the many varieties of entanglement. Our
focus here will be on understanding the communication resources
required to transform one known, bipartite, pure quantum state
into another. In the case where only local operations and
classical communication (LOCC) are allowed, we will find that
Klyachko's resolution \cite{Klyachko1} of Horn's Conjecture
\cite{Horn} gives an essentially complete answer, in principle, to
the question of which states can be converted into a given other
using a fixed finite amount of communication. The bulk of the
paper, however, is devoted to the case where qubits are exchanged
instead of classical bits, but that problem can again be resolved
in principle using a similar collection of techniques.

From a mathematical perspective, the first problem reduces via
well-known results to the question of determining the possible
eigenvalues of a convex combination of isospectral matrices with
defined spectra, a clear special case of Klyachko's result. We
also show how using only the much earlier Horn's Theorem, any such
convex combination can be replaced by a convex combination with
equal weights, an observation with implications for communication.
The second problem corresponds to determining the moment polytope
for the group $U(m)$ acting on $mn$ by $mn$ Hermitian matrices via
conjugation: $(X,H) \mapsto (X \ox I_n) H (X \ox I_n)^{-1}$. This
is a special case which we examine in detail of a problem
considered by Berenstein and Sjamaar \cite{BS98}. Again we find
simplifications not present in the general case.

\smallskip

\noindent {\bf Guide to the paper:} Section \ref{sec:Horn}
introduces Horn's problem and explains its relevance to state
transformations. Section \ref{sec:partial} then introduces the
partial trace problem which is the focus of the rest of the paper,
including a discussion of its physical interpretation. Section
\ref{sec:variational} presents a powerful variational approach to
sums of eigenvalues of a Hermitian matrix due to Hersch and
Zwahlen that is the source of our inequalities relating the
spectra of a matrix and its partial trace, while Section
\ref{sec:Schubert} consists entirely of background material about
the Schubert calculus. The heart of the paper is contained in
Sections \ref{sec:phi} and \ref{sec:inequalities} which provide,
respectively, the crucial cohomological calculation and an
explicit evaluation of the inequalities in low dimension. Also
included in section \ref{sec:inequalities} is a discussion of the
connection to representation theory and a very brief discussion of
the connection to symplectic geometry which is required to
complete our argument. (For more information on this aspect of the
problem, the reader can do no better than Knutson's excellent
review \cite{Knutson00}.)

\smallskip

\section{Horn's Problem and State Transformations}
\label{sec:Horn}

\subsection{The problem and its solution}
\label{subsec:HornProblem}

{\em Horn's problem} is the following:  Given the spectra of $n
\times n$ Hermitian matrices $X$ and $Y$, what are the possible
spectra of $Z = X + Y$?  This problem was first seriously attacked
by H.~Weyl in 1912 \cite{Weyl}, but the complete solution has only
been achieved recently \cite{Klyachko1, Klyachko2, Totaro, KT,
KTW, Belkale}.

Early attempts at Horn's problem involved finding inequalities
that the eigenvalues of $X$, $Y$, and $Z$ had to satisfy, in order
that $Z = X+Y$. Let $\a = (\a_1, \ldots, \a_n)$ be the eigenvalues
of $X$, $\b = (\b_1, \ldots, \b_n)$ be the eigenvalues of $Y$, and
$\g = (\g_1, \ldots, \g_n)$ be the eigenvalues of $Z$, all written
in non-increasing order. One basic constraint that $\a$, $\b$ and
$\g$ must satisfy is the {\em trace condition}
\begin{equation}
\label{eqn:trace} \sum_{i=1}^n \g_i = \sum_{i=1}^n \a_i +
\sum_{i=1}^n \b_i.
\end{equation}
Besides this equality condition, for many years all other known
constraints on the eigenvalues could be reduced to linear
inequalities among the eigenvalues; in fact, they all had the form
\begin{equation}
\label{eqn:eigsum} \sum_{k \in K} \g_k \leq \sum_{i \in I} \a_i +
\sum_{j \in J} \b_j.
\end{equation}
where $I$, $J$, and $K$ are all subsets of $\{1, \ldots, n\}$ of
the same cardinality $r$.  Such inequalities were systematically
analyzed by A.~Horn in 1962 \cite{Horn}.  He found conditions on
triples of index sets $(I, J, K)$ for which he conjectured that
inequalities of the form of Inequality~(\ref{eqn:eigsum}) would be
necessary and sufficient.

Horn defined sets $T_r^n$ of triples $(I, J, K)$, corresponding to
the (conjectured) necessary and sufficient inequalities
inductively as follows. For each positive integer $n$ and $r \leq
n$, let
\begin{equation}
U_r^n = \{(I, J, K) | \sum_{i \in I} i + \sum_{j \in J} j =
\sum_{k \in K} k + r(r+1)/2\}.
\end{equation}
Then for $r = 1$, let $T_1^n = U_1^n$.  For $r > 1$, let
\begin{eqnarray*}
T_r^n & = & \{(I, J, K) \in U_r^n | \mbox{ for all } p < r \mbox{ and all } (F, G,H )\in T_p^r, \\
& & \sum_{f \in F} i_f + \sum_{g \in G} j_g \leq \sum_{h \in H}
k_h + p(p+1)/2 \}.
\end{eqnarray*}
Horn then proposed:
\begin{conjecture}[Horn]
A triple $(\a, \b, \g)$ can be the eigenvalues of $n \times n$
Hermitian matrices $X$, $Y$, and $Z$, where $Z = X+Y$, if and only
if the trace condition holds, and
$$
\sum_{k \in K} \g_k \leq \sum_{i \in I} \a_i + \sum_{j \in J} \b_j
$$
for all $(I, J, K) \in T_r^n$, for all $r < n$.
\end{conjecture}

He showed that his conjecture was valid for $n = 3$ and $n = 4$
(the case $n =2$ was already known), and asserted that his proof
could be extended for $n \leq 8$. Moreover, he managed to prove
that the general form of his inequalities was sufficient:
\begin{theorem}[Horn] \label{thm:horn}
For each positive $n$ and $N$ there exists a finite set $L$ and
index sets $\{K_l\} \subset \{1, \ldots, N\}$ and $\{J_{il}\}
\subset \{1, \ldots, N\}$, where $l \in L$ and $i \in \{1, \ldots,
N\}$, such that the following holds:  An $n \times n$ Hermitian
matrix $A$ can be written as the sum of $N$ Hermitian $n \times n$
matrices with respective spectra $\l^1, \l^2, \ldots, \l^N$ if and
only if
\begin{equation}
\sum_{k\in K_l} (\Spec(A))_k \leq \sum_{i=1}^N \sum_{j \in J_{il}}
\l_j^i
\end{equation}
holds for all $l \in L$. (Note that the spectra $\l^i$ are each
written in non-increasing order.)
\end{theorem}
Extending the demonstration of the conjecture to the general case,
however, proved elusive. In 1982, B~.V.~Lidskii \cite{lidskii}
announced that he had verified Horn's conjecture, but his proof
sketch was very incomplete, and the details have never appeared.
The problem was finally definitively solved by Klyachko
\cite{Klyachko1, Klyachko2}, with important related contributions
from Belkale, Knutson, Tao, Totaro and Woodward \cite{Totaro, KT,
KTW, Belkale}:
\begin{theorem}
\label{thm:generalhorn} Horn's conjecture is true.
\end{theorem}

A somewhat unexpected complication that arises in that Horn's list
of inequalities is redundant for $n > 5$; as $n$ increases, the
number of redundant inequalities grows rapidly.  So it is natural
to desire a minimal set of inequalities that are necessary and
sufficient for $(\a, \b, \g)$ to be the spectra of Hermitian
matrices $X$, $Y$, and $X+Y$.  This issue has been resolved as
well; Knutson and Tao have developed combinatorial gadgets called
``honeycombs'' that can be used to determine which of Horn's
inequalities is redundant \cite{KT,KTW}.

\subsection{An Application to LOCC Protocols}
\label{subsec:HornLOCC}

Besides serving as a motivation for the present work, Horn's
problem itself yields insights into problems of quantum
information theory.  We present an application demonstrating that
it is sufficient to consider protocols of a special type in
performing transformations using LOCC.

Recall that given two probability vectors $p$ and $q$ in $\RR^N$,
we say that $q$ is majorized by $p$, or $q \prec p$, if
\begin{equation}
\sum_{i=1}^k q_i^\downarrow \leq \sum_{i=1}^k p_i^\downarrow
\end{equation}
for all $1 \leq k < N$, where $v^\downarrow$ represents the vector
$v$ with entries arranged in non-increasing order. There is a very
useful characterization of the majorization relation: $q \prec p$
if and only if $q = A p$ for some double stochastic matrix $A$.
The doubly stochastic matrices are, in turn, described by
\begin{theorem}[Birkhoff] \label{thm:Birkhoff}
The extreme points of the convex set of~~$n \times n$ doubly
stochastic matrices consist of~~$n \times n$ permutation matrices.
\end{theorem}
(See \cite{Bhatia} for a proof.) Given two Hermitian matrices $\r$
and $\s$, we write $\s \prec \r$ if $\Spec(\s) \prec \Spec(\r)$,
where $\Spec(X)$ is the vector of eigenvalues of an operator $X$.
(We will adopt the convention that $\Spec(X)$ is always written
with components in non-increasing order.) Uhlmann has proved a
matrix analog of Birkhoff's Theorem \cite{Uhlmann,AlbertiUhlmann}:
\begin{theorem}[Uhlmann]
Let $\r$ and $\s$ be~~$n \times n$ density matrices (that is,
positive semidefinite with unit trace). Then $\s \prec \r$ if and
only if there exists a probability vector $p$ and unitary matrices
$U_i$ such that
\begin{equation}
\s = \sum_i p_i U_i \r U_i^\dg.
\end{equation}
\end{theorem}
Using Theorem~\ref{thm:generalhorn}, we can strengthen Uhlmann's
Theorem.
\begin{theorem}
\label{thm:isospectral} Suppose a matrix $\s$ can be written as a
convex combination of isospectral Hermitian matrices:
\begin{equation}
\label{eqn:nonisosum} \s = \sum_{i=1}^N p_i U_i \r U_i^\dagger,
\end{equation}
where each $U_i$ is unitary, $p_i \geq 0$ and $\sum_i p_i = 1$.
Set $p = (p_1, \ldots, p_N)$, and suppose that $q$ is a
probability distribution such that $q \prec p$. Then there exist
unitary matrices $\{V_i\}_{i=1}^N$ such that
\begin{equation}
\s = \sum_{i=1}^N q_i V_i \r V_i^\dagger.
\end{equation}
\end{theorem}
{\bf Proof \/ }Let $\m = \Spec(\s)$ and $\l = \Spec(\r)$, so that
$p_i \l = \Spec(p_iU_i\r U_i^\dagger)$. By
Theorem~\ref{thm:generalhorn} there is a list of inequalities,
each of the form
\begin{equation}
\label{eqn:proofnoniso} \sum_{k \in K} \m_k \leq \sum_{i=1}^N
\sum_{J \in J_i} p_i \l_j,
\end{equation}
that must be satisfied in order for Equation~(\ref{eqn:nonisosum})
to hold.  By the symmetry of interchanging the order of the
summands in Equation~(\ref{eqn:nonisosum}), it must be true for
each $\p \in S_N$ that
\begin{equation}
\label{eqn:permsum} \sum_{k \in K} \m_k \leq \sum_{i=1}^N \sum_{j
\in J_i} p_{\p(i)} \l_j.
\end{equation}
Now since $q \prec p$, it follows from Theorem~\ref{thm:Birkhoff}
that there exist coefficients $c_\p \geq 0$, $\sum_{\p \in S_N}
c_\p = 1$, such that for all $i \in \{1, \ldots, N\}$,
\begin{equation}
\label{eqn:convexsum} q_i = \sum_{\p \in S_N} c_\p p_{\p(i)}.
\end{equation}
Now we take a convex sum of Inequalities~(\ref{eqn:permsum}) over
$\p \in S_N$:
\begin{eqnarray}
 \sum_{k \in K} \m_k & = & \sum_{\p \in S_N} c_\p \sum_{k \in K} \m_k \\
& \leq & \sum_{\p \in S_N} c_\p \sum_{i=1}^N \sum_{j \in J_i} p_{\p(i)} \l_j \mbox{ by Inequalities~(\ref{eqn:permsum})}\\
& = &  \sum_{i=1}^N \sum_{j \in J_i} \l_j \sum_{\p \in S_N} c_\p p_{\p(i)} \\
& =  & \sum_{i=1}^N \sum_{j \in J_i} q_i \l_j.
\end{eqnarray}
In other words, if an inequality of the form of
Inequality~(\ref{eqn:proofnoniso}) holds for values $p_i$, then it
also holds when every $p_i$ is replaced by $q_i$.  Applying
Theorem~\ref{thm:generalhorn}, we conclude that there must be
unitary matrices $\{V_i\}_{i=1}^N$ such that
\begin{equation}
\s = \sum_{i=1}^N q_i V_i \r V_i^\dagger.
\end{equation}
\hfill $\Box$

\noindent In particular, we have
\begin{corollary}
\label{cor:isospectral} Suppose a matrix $\s$ can be written as a
convex combination of unitary conjugations of a fixed Hermitian
matrix $\r$ with $N$ terms:
\begin{equation}
\s = \sum_{i=1}^N p_i U_i \r U_i^\dagger,
\end{equation}
where each $U_i$ is unitary, $p_i \geq 0$ and $\sum_i p_i = 1$.
Then there exist unitary matrices $\{V_i\}_{i=1}^N$ such that
\begin{equation}
\s = \frac{1}{N} \sum_{i=1}^N V_i \r V_i^\dagger.
\end{equation}
\end{corollary}
{\bf Proof \/} Set $q = (\frac{1}{N}, \ldots, \frac{1}{N})$ in
Theorem~\ref{thm:isospectral}. \hfill $\Box$

\smallskip

In \cite{NielsenLOCC}, M.~Nielsen described how to transform a
(pure) quantum state $|\ph_{AB}\rangle$, jointly held by two
parties, Alice and Bob, into another bipartite quantum state
$|\psi_{AB}\rangle$, using only local operations and classical
communication; he found that this is possible if and only if
\begin{equation}
\label{eqn:nielsenlocc} \Spec(\ph_A) \prec \Spec(\psi_A).
\end{equation}
(Note that L. Hardy independently arrived at the same conclusion
with the benefit of the benefit of Uhlmann's
Theorem~\cite{Hardy}.) It follows from Uhlmann's Theorem that
Condition~(\ref{eqn:nielsenlocc}) holds if and only if $\ph_A$ can
be written as a convex sum
\begin{equation}
\label{eqn:uhlmannrep} \ph_A = \sum_{i=1}^N p_i U_i \psi_A
U_i^\dagger
\end{equation}
where each $U_i$ is unitary. Nielsen showed that if
$|\ph_{AB}\rangle$ can be tranformed into $|\psi_{AB}\rangle$ via
LOCC, then Equation~(\ref{eqn:uhlmannrep}) holds, by presenting a
protocol (using $\log_2 N$ bits of classical communication) that
exhibits this representation.   In the protocol, one party
performs a measurement with $N$ possible outcomes, where $p_i$ is
the probability of the $i$th outcome, to her portion of the joint
system.  The outcome $i$ is communicated to the other party, who
then performs a unitary $U_i$ to his portion of the system. Any
such protocol carries out the transformation $|\ph_{AB}\rangle
\rightarrow |\psi_{AB}\rangle$.

In a subsequent paper, Harrow and Lo \cite{HarrowLo} demonstrated
that without altering the total number of bits transmitted, any
LOCC protocol for transforming known pure quantum states can be
transformed into one of the following form:
\begin{enumerate}
\item Alice performs a generalized measurement (POVM).
\item Alice sends the result of the measurement to Bob.
\item Bob performs a unitary operation conditioned on the message
he receives from Alice.
\item Alice and Bob both discard ancillary systems.
\end{enumerate}
With the exception of the discard step, the above protocol is of
the same type analyzed by Nielsen. Equation~(\ref{eqn:uhlmannrep})
therefore applies and Corollary~\ref{cor:isospectral} has the
following consequence.

\begin{corollary}
Any protocol for transforming known, pure, bipartite quantum
states via LOCC may be transformed into an equivalent one in which
all communication is from Alice to Bob, the total amount of
communication is the same \emph{and} all measurement outcomes, as
well as messages, are equiprobable.
\end{corollary}

\section{The Spectrum of a Partial Trace}
\label{sec:partial}

We now move on to defining our main problem. Let $A = {\mathbb
C}^{d_A}$, $B = {\mathbb C}^{d_B}$, and let $\rho _{AB}$ be an
operator on $A \otimes B$. We identify $\rho_{AB}$ with its matrix
in the standard basis, which has entries
\begin{equation}
\rho_{AB}^{ij,kl} = \langle i_A | \otimes \langle
j_B|\rho_{AB}|k_A\rangle \otimes |l_B\rangle
\end{equation}
in terms of orthonormal bases $\{\ket{i_A}\}$ and $\{\ket{j_B}\}$
of $A$ and $B$ respectively. Define the {\em partial trace}
$\rho_A
= \Tr_B \rho_{AB}$ of $\rho_{AB}$ to be the operator
\begin{equation}
\r_A = \sum_k \bra{k_B} \r_{AB} \ket{k_B}
\end{equation}
on $A$.  The matrix entries of $\r_A$ are
\begin{equation}
\r_A^{ij}
= \sum_k \bra{i_A} \ox \bra{k_B} \r_{AB} \ket{j_A} \ox \ket{k_B}.
\end{equation}
Equivalently, given the matrix $\r_{AB}$, we can define $\r_A$ to
be the unique matrix such that
\begin{equation} \label{eqn:invarDefn}
\Tr(\r_{AB} X \ox I_B) = \Tr(\r_A X)
\end{equation}
for all $X$ on $A$, where $I_B$ is the identity on $B$.

The rest of the paper will focus on the following question: What
is the relationship between the spectrum of $\rho_{AB}$ and the
spectrum of $\rho_A$? We generally adopt the point of view that
the spectrum of $\rho_{AB}$ is given and we wish to deduce which
possible spectra of $\rho_A$ may occur. (Our final results will
nonetheless allow us to reason in the other direction as well;
given the spectrum of $\rho_A$, one can deduce the possible
spectra of $\rho_{AB}$.)  We let $\cH_{AB}(\l) = \{ \r_{AB} :
\Spec(\r_{AB})
= \l \}$ be the set of Hermitian matrices on $A \ox B$ with
eigenvalues $\l_1 \geq \l_2 \geq \cdots \geq \l_{d_A d_B}$; then
our problem is to fully characterize the set $\cS_A(\l) = \{
\Spec(\r_A) : \r_{AB} \in \cH_{AB}(\l)\}$.

Some of the most useful inequalities in quantum information theory
relate the eigenvalues of a density matrix with those of its
partial traces. The von Neumann entropy $S(\r) = - \Tr \r \log_2
\r$, for example, satisfies inequalities known as subadditivity:
\begin{equation}
S(\r_{AB}) \leq S(\r_A) + S(\r_B)
\end{equation}
and, more generally, strong subadditivity~\cite{LiebRuskai}:
\begin{equation}
S(\r_{ABC}) + S(\r_B) \leq S(\r_{AB}) + S(\r_{BC}).
\end{equation}
Also, a good deal is known about the possible spectra of
reductions when the individual subsystems consist of qubits.
Bravyi \cite{Bravyi} and, independently Higuchi, Sudbery and Szulc
\cite{HSS}, for example, have found necessary and sufficient
conditions for a set of qubits states to be the reductions of a
larger pure state. In a similar spirit, Linden, Popescu and
Wootters \cite{LPW} have shown that almost all sets of states on
subsets of $n-1$ particles do not correspond to reductions of a
single $n$ particle state.

Most directly connected to the present paper, Christandl and
Mitchison very recently connected the triples of spectra of given
bipartite density matrix and its two reductions to the Kronecker
coefficients of the symmetric group~\cite{CM}. We make a similar
connection in Section \ref{subsec:repnTheory}, providing a
correspondence of the type linking Horn's Problem to the
Littlewood-Richardson coefficients. \footnote{Coincidentally, days
after the initial submission of this manuscript, Klyachko also
published a very insightful paper on the same
topic~\cite{Klyachko04}.}

\subsection{Physical Interpretation} \label{subsec:partialInterp}

Determining the possible spectra of a partial trace has a number
of physical applications. The usual situation is to regard
$\rho_{AB}$ as the density matrix of a quantum system $AB$, a
composite of subsystems $A$ and $B$; $\rho_A$ is then the density
matrix of subsystem $A$. In this context, we are asking which
quantum-mechanical descriptions of a subsystem of a quantum system
are compatible with the description of the whole system.

Understanding the relationship between a density operator and its
partial trace also allows us to characterize which state
transformations are achievable using quantum communication.  To
illustrate, suppose two parties, Alice and Bob, share a state
between them that can be described by a state vector
$|\ph_{ABC}\rangle \in A \ox B \ox C$, where Alice holds quantum
systems $A$ and $C$ and Bob holds the system $B$.  First, we
assume that there will be only one round of quantum communication,
from Alice to Bob.   Alice's initial description of her subsystem
(her \emph{reduced density operator}) is given by $\ph_{AC} =
\Tr_B |\ph_{ACB}\rangle\langle \ph_{ACB}|$.  If she then sends Bob
the system $C$ through a \emph{quantum channel}, her new density
operator becomes $\ph_A = \Tr_{CB}|\ph_{ACB}\rangle \langle
\ph_{ACB}|$. Thus, understanding how quantum systems change as a
result of quantum communication is equivalent to understanding how
a density matrix is related to its partial trace. (It is important
to note, however, that the model is slightly different than in the
LOCC case. Instead of allowing arbitrary local operations, we
allow only unitary local operations.)

\begin{figure}[t]
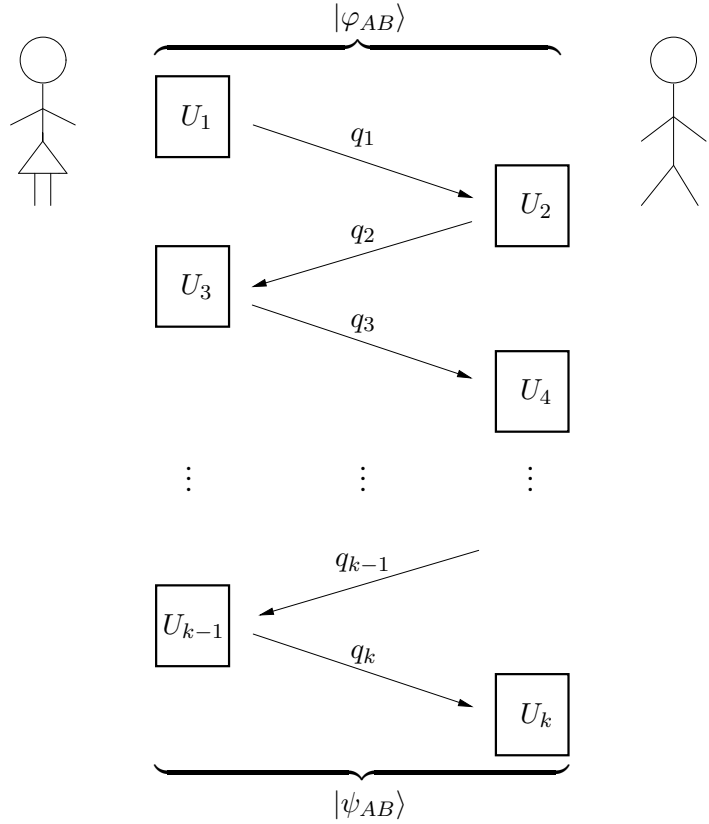

\begin{center}
\input loqc.eepic
\end{center}
\caption[A many-round quantum communication
protocol]{\label{fig:loqc} A many-round quantum communication
protocol.  Two parties, Alice and Bob, initially share a joint
system $|\ph_{AB}\rangle$.  Alice applies a local unitary operator
$U_1$ and then sends $q_1$ quantum bits to Bob, who performs a
local unitary $U_2$ and then sends $q_2$ quantum bits to Alice,
etc.; in the end, they share system $\psi_{AB}\rangle$.
  By Theorem~\ref{thm:oneround}, this is equivalent to a protocol in which there
is only one round of quantum communication.}
\end{figure}

If many rounds of communication are allowed in a quantum
communication protocol, it may seem that the analysis should
become more complicated (see Figure~\ref{fig:loqc}).
Happily, this turns out not to be the case.  In fact, the following
result \cite{vDH02} shows that it is enough to consider one-round protocols:
\begin{theorem}
\label{thm:oneround}
Suppose there exists a bipartite quantum communication protocol that transforms the state
$|\varphi_{AB}\rangle$ to the state $|\psi_{AB}\rangle$, requiring a total of $q$ qubits of communication.
Then there is a one-round protocol that accomplishes the same transformation
$|\varphi_{AB}\rangle \rightarrow |\psi_{AB}\rangle$, also requiring $q$ qubits of information.
\end{theorem}
{\bf Proof \/ }The proof involves showing that at any round of the
protocol, any communication from Bob to Alice can be replaced by
communication from Alice to Bob; it then follows that all
communication can be taken to be in one direction.  The effect of
Bob sending a qubit to Alice is to transform a state $\sum_i
\sqrt{\l_i} \ket{i_A}\ket{i_B}$ to a state $\sum_i
\sqrt{\l_{i'}}\ket{i_A'}\ket{i_B'}$, where the prior and posterior
states are written in their Schmidt decompositions. (By an
application of the singular value decomposition, any bipartite,
quantum state $\ket{\o}$ is equivalent by local unitary
transformations to one of the form $\ket{\o'} = \sum_i \sqrt{\a_i}
\ket{i_A} \ket{i_B}$, where $\sum_i \a_i = 1$ and
$\braket{i_A}{j_A} = \braket{i_B}{j_B} = \d_{ij}$. This is known
as the Schmidt decomposition.) But by symmetry, we then see that
the swap operator exchanging Alice's and Bob's systems is
equivalent to applying some local unitaries $U_A \ox U_B$ on their
joint system. Thus, instead of having Bob send a qubit to Alice,
they can apply $U_A \ox U_B$ and then have Alice send a qubit to
Bob (and finally apply some local unitaries $U_A' \ox U_B'$ to
swap Alice and Bob back again) to accomplish the same
transformation. \hfill $\Box$

While the problem of comparing the spectrum of a matrix to that of its partial trace has a natural
application to density matrices, it may be applied to other settings as well.
For example, given the spectrum of an observable for a certain quantum system,
one may wish to ask what the spectrum of
that observable may be
for a subsystem of the given system.  In this context $\r_{AB}$ is the matrix
of the observable, rather than a density matrix.

\section{Variational Principle} \label{sec:variational}
\label{ch:varprin}

We use a variational principle argument to show that inequalities
between the eigenvalues of $\r_{AB}$ and of $\r_A$ arise whenever
certain subsets of the Grassmannian intersect. We also show
explicitly that when $d_A = 2$, these inequalities are sufficient.

\subsection{Some Basic Inequalities}
\label{sec:kyfan}

In this section we use a simple argument to derive some inequalities that the spectra of $\r_{AB}$ and
$\r_A$ must satisfy.  Although these inequalities will subsumed by our later results, the proof
illustrates the strategy behind the general method.
We will make use of the following well-known fact from linear algebra~\cite{Bhatia}:

\begin{lemma}[Ky Fan's Maximum Principle]
\label{lem:kyfan} Let $A$ be an $n \times n$ Hermitian matrix with
spectrum $\l$, where we assume as usual that the components of
$\l$ are in non-increasing order.  Then for all $k \in \{1,
\ldots, n\}$,
\begin{equation}
\label{eq:fan} \sum_{j=1}^k \l_j = \max \sum_{j=1}^k \bra{x_j} A
\ket{x_j}
\end{equation}
where
the maximum is taken over all orthonormal $k$-tuples of vectors
$\{\ket{x_j}\}_{j=1}^k$ in ${\mathbb C}^n$.
\end{lemma}

If $V$ is a subspace of the
vector space $A$, then we write $V \leq A$. Thus, $\mbox{Gr}_k(A)
:= \{ V \leq A: \dim(V)=k \}$ is the Grassmannian of
$k$-dimensional subspaces of the vector space $A$. We also write
$\Gr(k,n)$ for $\Gr_k(\CC^n)$. For $V \leq \CC^n$ with the
standard inner product, let $P_V$ denote the orthogonal projection
operator onto the subspace $V$. Given a vector $v \in \CC^d$ and a
positive integer $n$, we define $\Sigma_n(v)$ to be the vector
whose components are obtained by summing successive blocks of $n$
components of $v$:
\begin{equation}
\Sigma_n(v) = (v_1 + \cdots + v_n, v_{n+1} + \cdots + v_{2n},
\ldots, v_{\lfloor d/n \rfloor (n-1) + 1} + \cdots + v_d).
\end{equation}
Recall that we denoted the dimensions of system $A$ and $B$ by $d_A$ and $d_B$, respectively; and that all
vectors of matrix spectra are assumed to be with components in non-increasing order.  We will use these
conventions throughout.

\begin{lemma}
\label{lem:basicineq} Let $\l$ be the spectrum of $\r_{AB}$, and
$\tilde \l$ be the spectrum of its partial trace $\r_A$. Then for
every $k \in \{1, \ldots, d_A\}$, the inequality
\begin{equation}
\sum_{i=1}^k \tilde \l_i \leq \sum_{i=1}^{d_B k} \l_i
\end{equation}
must hold.
We may write the $d_A$ inequalities succinctly as the majorization relation
\begin{equation}
\tilde \l \prec \Sigma_{d_B}(\l).
\end{equation}
\end{lemma}
{\bf Proof \/ }\begin{equation}
\label{eqn:redundant}
\begin{array}{rcl}
\displaystyle{\sum_{i=1}^k \tilde \l_i} & = & \max_{\{V \in \mbox{ Gr}_k(A)\}} \Tr(\r_A P_V)\\
 & = &  \max_{\{V \in \mbox{ Gr}_k(A)\}} \Tr(\r_{AB} P_{V \ox B})\\
 & \leq &  \max_{\{V \in \mbox{ Gr}_{kd_B}(A \ox B)\}} \Tr(\r_{AB} P_{V})\\
 & = & \displaystyle{\sum_{i=1}^{kd_B} \l_i},
\end{array}
\end{equation}
where the first and last equalities follow from Ky Fan's Maximum Principle, the
second equality comes from the definition of partial trace, and the inequality follows
because the maximum is being taken over a larger set of projection operators than in the
previous expression.
\hfill $\Box$

Note the basic idea behind the proof.  We expressed the sum of eigenvalues for each matrix
in terms of a
variational principle on subspaces, and then we looked for an intersection between subspaces
in order to relate the variational expressions. This idea will be developed further in the
next section.

\subsection{General Method}

Let $A$ be an $n \times n$ Hermitian matrix with spectrum $\l$,
and let $V$ be a subspace of $\CC^n$. Define the {\em Rayleigh
trace} of $A$ on $V$ to be
\begin{equation}
R_A(V) = \Tr( P_V A ).
\end{equation}
Observe that if $B$ is another Hermitian matrix, then $R_{A+B}(V)
= R_A(V) + R_B(V)$. We can restate Ky Fan's Maximum Principle in
this notation:
\begin{equation}
\underset{V, \mbox{ dim }V = r}{\max} R_A(V) = \sum_{i=1}^r \l_i.
\end{equation}
Likewise,
\begin{equation}
\min_{V, \mbox{ dim }V = r} R_A(V) = \sum_{i=n-r+1}^n \l_i.
\end{equation}
Let $A_r$ denote the $r$-dimensional vector space spanned by eigenvectors corresponding to the $r$
largest eigenvalues of $A$ (if $A$ is degenerate with $\l_r$ = $\l_{r+1}$, then choose any such $A_r$.)
Now given a binary sequence $\p$ of length $n$ and weight $r$ (sometimes written
$\p \in {n \choose r}$), the
{\em Schubert cell} in the $r$-Grassmannian corresponding to $\p$ is defined as
\begin{equation}
S_\p(A) = \{V \leq \CC^n| \mbox{ dim}(V\cap A_i)/(V \cap A_{i-1}) = \p(i), 1 \leq i \leq n\},
\end{equation}
where $\p(i)$ is the $i$th term in the sequence $\p$. Then $\p(i)
= 1$ for $r$ values of $i$; label these values $i_1 < i_2 < \cdots
i_r$. The following variational principle, due to Hersch and
Zwahlen~\cite{HZ62}, provides access to sums of arbitrary
combinations of the eigenvalues of $A$:
\begin{theorem}
\label{thm:HZ}
\begin{equation}
\min_{V \in S_\p(A)} R_A(V) = \sum_i \p(i)\l_i.
\end{equation}
Equality occurs when $V$ is the span of eigenvectors corresponding to the eigenvalues $\l_{i_1}, \ldots, \l_{i_r}$.
\end{theorem}
{\bf Proof \/ } Let $V \in S_\p(A)$, and choose orthogonal unit
vectors $\ket{u_1}, \ket{u_2}, \ldots, \ket{u_r}$ such that
$\ket{u_k} \in V \cap A_{i_k}$. Now $A_{i_k}$ is spanned by
eigenvectors of $A$ with eigenvalue greater than or equal to
$\l_{i_k}$, so $\bra{u_k} A \ket{u_k} \geq \l_{i_k}$. It follows
that
\begin{equation}
R_A(V) = \sum_{k=1}^r \bra{u_k} A \ket{u_k} \geq \sum_{k=1}^r
\l_{i_k} = \sum_i \p(i) \l_i.
\end{equation}
Now suppose $V$ is the span of eigenvectors corresponding to eigenvalues $\l_{i_1}, \l_{i_r}$.
In this case $u_k$ is an eigenvector of $A$ with eigenvalue $\l_{i_k}$, so that $R_A(V) = \sum_i \p(i) \l_i$.
\hfill $\Box$

\smallskip

Now let $B = \CC^{d_B}$. For any $k \leq d_A$, define the map
$\phi: \Gr_k(A) \rightarrow \Gr_{d_Bk}(A \otimes B)$ by $\phi(V) =
V \otimes B$. Let $\ket{y_1}, \ldots, \ket{y_{d_B}}$ be an
orthonormal basis of $B$, and let $I_B$ denote the identity
operator on $B$. For any operator $X_A$ on $A$, and any $\ket{v}
\in \CC^{d_A}$, we have that
\begin{eqnarray*}
\sum_{i=1}^{d_B}\bra{v} \ox \bra{y_i} \Big( \frac{1}{d_B} X_A \ox
I_B \Big) \ket{v} \ox \ket{y_i} & = &
\frac{1}{d_B}\sum_{i=1}^{d_B}\bra{v} X_A \ket{v}
\bra{y_i} I_B \ket{y_i} \\
& = & \frac{1}{d_B} \sum_{i=1}^{d_B} \bra{v} X_A \ket{v} \\
& = & \bra{v} X_A \ket{v}.
\end{eqnarray*}
It follows that $R_{X_A}(V) = R_{\frac{1}{d_B} X_A \ox I_B}(\phi(V))$.

The following theorem was motivated by an analogous argument, due
to Johnson \cite{Johnson79} and the pair of Helmke and Rosenthal
\cite{HR95}, used in the solution of Horn's problem.

\begin{theorem}
\label{thm:varprin}
Let $X_A$ be an operator on $A$ and $Y_{AB}$ be an operator on $A \ox B$ such that
$X_A = -\Tr_B(Y_{AB})$.
Let $\tilde \l$ be the spectrum of $X_A$ and $\l$ be the
spectrum of $Y_{AB}$.
If $\phi(S_\p(X_A)) \cap S_\s(Y_{AB}) \not = \emptyset$, then
\begin{equation}
\label{eqn:varprin}
\sum_{i=1}^{d_A} \p(i)\tilde \l_i + \sum_{i=1}^{d_A d_B} \s(i)\l_i \leq 0.
\end{equation}
Inequality~(\ref{eqn:varprin}) also holds if
$\phi(\overline{S_\p(X_A)}) \cap \overline{S_\s(Y_{AB})} \not =
\emptyset$.
\end{theorem}
{\bf Proof \/ }
Let $W \ox B \in  \phi(S_\p(X_A)) \cap S_\s(Y_{AB})$.  Then we have
\begin{eqnarray}
& \sum_{i=1}^{d_A} \p(i)\tilde \l_i + \sum_{i=1}^{d_A d_B} \s(i)\l_i  \nonumber \\
= & \displaystyle{\min_{V \in S\p(X_A)}R_{X_A}(V) +
\min_{V' \in S_\s(Y_{AB})}R_{Y_{AB}}(Y_{AB})} \nonumber \\
 = & \displaystyle{\min_{V \in \phi(S\p(X_A))}R_{\frac{1}{d_B}X_A \ox I_B}(V) +
\min_{V' \in S_\s(Y_{AB})}R_{Y_{AB}}(Y_{AB})} \nonumber \\
 \leq & R_{\frac{1}{d_B}X_A \ox I_B}(W \ox B) + R_{Y_{AB}}(W \ox B) \nonumber \\
 = & R_{\frac{1}{d_B}X_A \ox I_B + Y_{AB}}(W \ox B) \nonumber \\
= & \Tr(P_W (X_A + \Tr_B(Y_{AB}))) \nonumber \\
= & 0.
\end{eqnarray}
This proves the inequality in the case that  $\phi(S_\p(X_A)) \cap
S_\s(Y_{AB}) \not = \emptyset$. If $\phi(\overline{S_\p(X_A)})
\cap \overline{S_\s(Y_{AB})} \not = \emptyset$, then
Theorem~\ref{thm:HZ}, along with the fact that the Rayleigh trace
is continuous, implies that $\min_{V \in \overline{S_\p(A)}}
R_A(V) = \sum_i \p(i)\l_i$, and the argument for the case
$\phi(S_\p(X_A)) \cap S_\s(Y_{AB}) \not = \emptyset$ applies
equally to this case. \hfill $\Box$

Starting from intersections of Schubert cells,
Theorem~\ref{thm:varprin} yields inequalities that must be
satisfied by the spectra of a matrix and its partial trace. As we
will discuss in the next section, the closures of the Schubert
cells are generators of the homology of the Grassmannian; thus, we
can regard the inequalities as coming from nonzero products in
cohomology. Determining which of these products are nonzero and
translating these nonzero products into the appropriate
inequalities will be the focus of the remainder of the paper.

\subsection{Solution for $d_A = 2$}
\label{sec:dim2sol}

When $d_A = 2$, the relationship between the spectrum of $\r_{AB}$
and that of $\Tr(\r_{AB}) = \r_A$ is particularly simple: the only
inequalities restricting the spectra are those given by
Lemma~\ref{lem:basicineq}.  Moreover, in this case it is possible
to give a very simple and explicit construction of matrices
demonstrating that the inequalities are sufficient. (If we
interpret our problem in terms of quantum communication protocols,
the $d_A = 2$ case corresponds to the situation where Alice sends
to Bob her entire quantum system except for one qubit.)

\begin{theorem}
\label{thm:dim2case} If $d_A = 2$, the inequalities given by
Lemma~\ref{lem:basicineq} are sufficient. That is, given a vector
$\l \in \RR^{2d_B}$ and a vector $\tilde \l \in \RR^2$, each with
components in non-increasing order, satisfying $\tilde \l \prec
(\sum_{i=1}^{d_B} \l_i, \sum_{i=d_B + 1}^{2d_B} \l_i)$, there
exist matrices $\r_{AB}$ and $\r_A$ such that the spectrum of
$\r_{AB}$ is $\l$, the spectrum of $\r_A$ is $\tilde \l$, and
$\r_A = \Tr_B(\r_{AB})$.
\end{theorem}
{\bf Proof \/ } Let $\l =
(\l_{0,0}, \l_{0,1}, \ldots, \l_{0, d_B -1}, \l_{1, 0}, \l_{1, 1}, \ldots, \l_{1, d_B - 1}$),
let $\{|0_A\rangle, |1_A\rangle \}$ and $\{|0_B\rangle, \ldots, |(j-1)_B\rangle$ be orthonormal
bases for $A$ and $B$, respectively, and
set
\begin{equation}
\s_{AB} = \sum_{i=0}^1 \sum_{j=0}^{d_B-1} \l_{i, j} |i_A\rangle |j_B \rangle \langle i_A|\langle j_B|
\end{equation}
For $t \in [0, 2\p]$, let
\begin{equation}
\begin{array}{cl}
U(t) = & \displaystyle{\sum_{i=0}^1 \sum_{j=0}^{d_B-1} \cos t |i_A\rangle |j_B \rangle \langle i_A|\langle j_B|}\\
& + \displaystyle{\sum_{j=0}^{d_B-1} \sin t |0_A\rangle |(j-1)_B\rangle \langle 1_A| \langle j_B|}\\
& - \displaystyle{\sum_{j=0}^{d_B-1} \sin t |1_A\rangle |j_B\rangle \langle 0_A| \langle (j-1)_B|},
\end{array}
\end{equation}
where the subtraction in the labels of the bra and ket vectors is done modulo $d_B$.
Now $U(t)$ is unitary (in fact, it is real orthogonal) for all $t$, so the
spectrum of $U(t)\s_{AB}U(t)^\dagger$ is $\l$. A direct calculation verifies that
\begin{equation}
\begin{array}{cl}
U(t) \s_{AB} U(t)^\dagger = &
 \displaystyle{\sum_{i=0}^1 \sum_{j=0}^{d_B-1} \l_{i,j} \cos^2 t |i_A\rangle |j_B \rangle \langle i_A|\langle j_B|}\\
 & + \displaystyle{\sum_{j=0}^{d_B-1} (\l_{1,j}-\l_{0,j-1})\sin t \cos t |0_A\rangle |(j-1)_B\rangle \langle 1_A| \langle j_B|}\\
 & + \displaystyle{\sum_{j=0}^{d_B-1} (\l_{1,j}-\l_{0,j-1})\sin t \cos t |1_A\rangle |j_B\rangle \langle 0_A| \langle (j-1)_B|}\\
 & +  \displaystyle{\sum_{j=0}^{d_B-1} \sin^2 t (\l_{0,j-1} |0_A\rangle |j_B\rangle \langle 0_A| \langle j_B| +
 \l_{1,j} |1_A\rangle |j_B\rangle \langle 1_A| \langle j_B|)},
\end{array}
\end{equation}
so that
\begin{equation}
\begin{array}{cl}
\Tr_B(U(t)\s_{AB}U(t)^\dagger) =
& \big( \sum_{j=0}^{d_B -1} \l_{0,j}\cos^2 t + \sum_{j=0}^{d_B -1}\l_{1,j} \sin^2 t \big) |0_A\rangle \langle 0_A|\\
 & + \big( \sum_{j=0}^{d_B -1} \l_{1,j}\cos^t + \sum_{j=0}^{d_B -1}\l_{0,j} \sin^2 t \big) |1_A\rangle \langle 1_A|.
\end{array}
\end{equation}

Let $\a_1 = \sum_{j=0}^{d_B -1} \l_{0,j}$, $\a_2 =  \sum_{j=0}^{d_B -1} \l_{1,j}$.
If we let $\r_{AB}(t) = U(t)\s_{AB}U(t)^\dagger$, then the spectrum of the partial trace of
$\r_{AB}(t)$ is $(\a_1 \cos^2 t + \a_2 \sin^2 t, \a_1 \sin^2 t + \a_2 \cos^2 t)$.  By
choosing the appropriate value of $t \in [0, 2\p]$, any convex combination of $\a_1$
and $\a_2$ can be achieved for the eigenvalues of $\Tr_B(\r_{AB}(t))$.
\hfill $\Box$

\section{Schubert Calculus} \label{sec:Schubert}

This section is intended as a quick introduction to arithmetic in
the cohomology ring of the Grassmannian, otherwise known as the
Schubert calculus. While we include some proofs in order to try to
help the reader understand the nature of the arguments, our
presentation is necessarily incomplete. Full treatments, upon
which the following discussion is based, can be found in
\cite{Manivel01}, \cite{Fulton97}, and \cite{GH78}.

\subsection{Symmetric Polynomials}

We start with some background on the ring $\L_n$ of symmetric
polynomials in $n$ variables with integer coefficients. A certain
class of such polynomials, the Schur polynomials, will be of
particular interest, due to its relationship with the cohomology
of the Grassmannian.  The Schur polynomials (as well as the
Grassmannian cohomology classes) are indexed by partitions of
integers, so we begin with some terminology relating to
partitions.

A {\em partition} of an integer $n$ is a finite sequence $\a = (\a_1, \ldots \a_l)$ of nonnegative
integers, with $n = \sum_i \a_i$, arranged in non-increasing order: $\a_1 \geq \a_2 \geq \cdots \geq \a_l \geq 0$.  These integers
$\a_1, \ldots, \a_l$ are called the {\em parts}, and the {\em length} $\ell(\a)$ is the number of
nonzero parts.  The integer $n = \sum_i \a_i$ is the {\em weight} of the partition, denoted $|\a|$.
To any partition $\a$ we may associate a Young diagram, whose $i$th row has length $\a_i$.  The
{\em conjugate partition}
 $\a^*$ is obtained by interchanging rows and columns in the Young diagram of $\a$.  For instance,
if $\a = (5, 3, 2, 2)$, then the Young diagram of $\a$ is
\begin{equation}
\yng(5,3,2,2)
\end{equation}
so the Young diagram of $\a^*$ is
\begin{equation}
\yng(4,4,2,1,1)
\end{equation}
and $\a^* = (4, 4, 2, 1, 1)$.

Now let $\L_n$ be the ring of symmetric polynomials with integer
coefficients in $n$ variables. There are a number of
computationally useful bases for $\L_n$.  Perhaps the simplest
basis is given by the {\em monomial symmetric functions.} These
are functions obtained by starting with a monomial $x^\a =
x_1^{\a_1} \cdots x_n^{\a_n}$ and symmetrizing it, to obtain a
polynomial
\begin{equation}
m_\a = \sum_{\b \in S_n(\a)} x^\b.
\end{equation}
In this notation, $S_n$ permutes the coefficients of $\a$.  Note that the sum is not over all permutations in $S_n$, but
over the image of these permutations; thus, any given monomial appears only once in the sum.
\begin{theorem}
\label{thm:monomial}
The polynomials $m_\a$, where $\a$ ranges over partitions with at most $n$ parts, form a basis over $\ZZ$ for
the ring $\L_n$.
\end{theorem}
{\bf Proof \/ } Given a polynomial $p(x_1, \ldots, x_n) = \sum
c_\a x^\a \in \L_n$, let $\a = (\a_1, \ldots, \a_n)$ be the
maximal $n$-tuple (with respect to the lexicographic ordering)
such that $c_\a \not = 0$.  Because $p(x_1, \ldots, x_n)$ is
symmetric, $\a$ must be a partition. Now $p(x_1, \ldots, x_n) -
c_\a m_\a$ is also a symmetric polynomial, but one whose leading
monomial is smaller than $x^\a$ with respect to the lexicographic
ordering.  Because $\a_i \geq 0$, the lexicographic ordering is a
well-ordering, so it follows by induction that $p(x_1, \ldots,
x_n)$ can be written as an integer combination of terms $m_\a$.

Now suppose $\sum c_\a m_\a = 0$.  Again, let $\a$ be the maximal $n$-tuple with respect to the lexicographic ordering
such that $c_\a \not = 0$.  Then the coefficient of $x^\a$ in the polynomial $\sum c_\a m_\a$ is $c_\a$, a contradiction.
\hfill $\Box$

We will make reference to the following two classes of symmetric polynomials.
The {\em elementary symmetric polynomials} are a subset of the monomial symmetric functions, corresponding
to partitions such that all parts are equal to one:
\begin{equation}
e_k = \sum_{1 \leq i_1 < \cdots < i_k \leq n} x_{i_1}\cdots x_{i_k},
\end{equation}
for $1 \leq k \leq n$.  The {\em complete symmetric polynomials} are
\begin{equation}
h_k = \sum_{1 \leq i_i \cdots \leq i_k \leq n} x_{i_1} \cdots x_{i_k},
\end{equation}
for $1 \leq k \leq n$.  (If $k = 0$, then set $e_0 = h_0 = 1$.)
We label products of elementary symmetric polynomials, as well as products
of complete symmetric polynomials, by partitions $\a$: $e_\a = e_{\a_1}\cdots e_{\a_l}$, and
$h_\a = h_{\a_1}\cdots h_{\a_l}$.

Both the elementary symmetric polynomials and complete symmetric polynomials
are important objects in the study of the ring $\L_n$.  The {\em fundamental theorem of symmetric polynomials}
states that every symmetric polynomial can be written as a polynomial in the elementary symmetric
polynomials \cite{Edwards84}; in other words, the polynomials $e_\a$, where $\a$ ranges through partitions with parts
less than or equal to $n$, form a basis over $\ZZ$ of the ring $\L_n$.    We will make use of the following
relationship between the polynomials $e_k$ and $h_k$.

\begin{proposition}
\label{thm:involution}
Let $\o:\L_n \rightarrow L_n$ be the ring homomorphism defined by $\o(e_k) = h_k$.  Then $\o$ is an
involution.
\end{proposition}

It follows from the fundamental theorem of elementary symmetric polynomials and
Proposition~\ref{thm:involution} that the polynomials $h_\a$
form a $\ZZ$-basis of $\L_n$.
We now describe another basis for the ring $\L_n$: the {\em Schur polynomials}, which will be a greater focus of our
study.  In order to do so, we make some
observations about the ring of antisymmetric polynomials in $n$ variables.  These polynomials have a basis
obtained from antisymmetrizing monomials: if $\g$ is an $n$-tuple of natural numbers, then let
\begin{equation}
a_\g = \sum_{w \in S_n} \varepsilon(w) x^{w(\g)},
\end{equation}
where $\varepsilon(w)$ is the sign of the permutation $w$.  Note that if $\g$ has two equal components, then
$a_\g = 0$.  Thus, we restrict our attention to the case where $\g$ is a strictly decreasing partition.  Then $\g$ has the
form
$\g = \a + \d$, where $\a$ is a partition and $\d = (n-1, n-2, \ldots, 1, 0)$.  An argument similar to the proof of
Theorem~\ref{thm:monomial} shows that the polynomials $a_{\a + \d}$, where $\a$ ranges over partitions with
at most $n$ parts, form a basis for the ring of antisymmetric polynomials with integer coefficients.

Next, note that every antisymmetric polynomial must be divisible by $(x_i - x_j)$ for all $i \not = j$, and so must
be divisible by the Vandermonde determinant $\det(x_i^{n-j})_{1\leq i, j \leq n} = \prod_{1\leq i < j \leq n} (x_i - x_j)$.
It is not hard to see that multiplying a symmetric polynomial by the Vandermonde determinant produces an
antisymmetric polynomial, and that dividing an antisymmetric polynomial by the Vandermonde determinant yields
a symmetric polynomial.  Thus, multiplication by the Vandermonde determinant gives an isomorphism between symmetric
and antisymmetric polynomials.  The {\em Schur polynomials} are obtained by dividing the polynomials $a_\g$ by the
Vandermonde determinant (which is the same as $a_\d$):
\begin{equation}
s_\a = \frac{a_{\a + \d}}{a_\d} = \frac{\det(x_i^{\a_j + n - j})_{1 \leq i,j \leq n}}{\det (x_i^{n-j})_{1\leq i, j \leq n}}.
\end{equation}
By the isomorphism between symmetric and antisymmetric polynomials, we have proven the following theorem.

\begin{theorem}
\label{thm:schurbasis}
The Schur polynomials $s_\a$, as $\a$ ranges over all
partitions with at most $n$ parts, form a basis over $\ZZ$ of the ring $\L_n$.
\end{theorem}

Given a partition $\a$ and integer $k$, let $\a \ox k$ denote the set of partitions obtained by adding $k$ boxes
to (the Young diagram of) $\a$, at most one box per column.  Let $\a \ox 1^k$ denote the set of partitions obtained
by adding $k$ boxes to $\a$, at most one box per row.

\begin{theorem}[Pieri formulas]
\label{thm:pieri}
With the above notation,
\begin{equation}
s_\a e_k = \sum_{\b \in \a \ox 1^k} s_\b,
\end{equation}
and
\begin{equation}
s_\a h_k = \sum_{\b \in \a \ox k} s_\b.
\end{equation}
\end{theorem}

The Pieri formulas, for example, can be used to write the Schur
polynomials in terms of the complete symmetric polynomials:
\begin{theorem}[Jacobi-Trudi formula]
Let $\a$ be a partition with at most $n$ parts. Then
\begin{equation}
s_\a = \det(h_{\a_i - i + j})_{1 \leq i, j \leq n}.
\end{equation}
\end{theorem}
{\bf Proof \/ }
Let $l$ be the length of $\a$.  Because $h_0 = 1$, $\det(h_{\a_i - i + j})_{1 \leq i, j \leq n} =
\det(h_{\a_i - i + j})_{1 \leq i, j \leq l}$.  Expand $\det(h_{\a_i - i + j})_{1 \leq i, j \leq l}$
along the last column, using induction on $l$:
\begin{equation}
\label{eqn:jactrusum}
\det(h_{\a_i - i + j})_{1 \leq i, j \leq l} = \sum_{i=1}^l (-1)^{l - i} s_{\l_1, \ldots, \l_{i-1}, \l_{i+1}-1,
\ldots, \l_l -1} \times h_{\l_i+ l -i}.
\end{equation}
Now it follows from Theorem~\ref{thm:pieri} that the $i$th term of the above sum may be written as
\begin{equation}
\sum_{\b \in J_i} s_\b + \sum_{\b \in J_{i+1}} s_\b,
\end{equation}
where $J_i$ is the set of partitions $\b$ having the same weight as $\a$, satisfying the conditions
$\a_j \leq \b_j \leq \a_{j-1}$ for $j < i$, and $\a_{j+1}-1 \leq \b_j \leq \a_j - 1$ for $j \geq i$.
Therefore, the right hand sum of Equation~\ref{eqn:jactrusum} telescopes to give us the desired formula.
\hfill $\Box$

\subsection{Grassmannians as Varieties}

Let $E$ be an $n$-dimensional complex vector space.
The Grassmannian $\Gr(k,n)$ can be realized as the homogeneous
space $U(n) / \big( U(k) \times U(n-k) \big)$ since the larger
group acts transitively on subspaces of $\CC^n$ while the smaller
one is the stabilizer of a fixed subspace. $\Gr(k,n)$ is, in fact,
a complex manifold of dimension $k(n-k)$.

If $V$ is a $k$-dimensional subspace of $E$, then $\wedge^k V$ is a line in $\wedge^k E$, giving us a map
\begin{equation}
\phi: \Gr_k(E) \rightarrow \PP(\wedge^k E),
\end{equation}
where we have introduced the notation $\PP(V)$ for the
projectivization of the vector space $V$. Let $A
= (a_{ij})$ be a $k \times n$ matrix representing $V$, so that $V$
is the span of the rows of $A$. Then a set of homogeneous
coordinates in $\phi(V)$ is given by the determinants of the $k
\times k$ minors of this matrix: if $I$ is a subset of $\{1,
\ldots, n\}$ of cardinality $k$, then define the coordinate
\begin{equation}
x_I = \det{A_I},
\end{equation}
where $A_I$ denotes the $I$th $k \times k$ minor of $A$.  These
coordinates are known as {\em Pl\"ucker coordinates}, and the map
$\phi$ is called the {\em Pl\"ucker embedding}. It can be shown
\cite{Manivel01} that the Pl\"ucker embedding is indeed an
embedding of the Grassmannian $\Gr_k(E)$ into the projective space
$\PP(\wedge^k E)$, and that the homogeneous coordinates are the
solutions of
 a set of (quadratic) polynomial equations, giving $\Gr_k(E)$ the structure of a projective algebraic variety.

\subsection{Schubert Varieties}

Define a {\em (complete) flag} $F_\bullet$ on $E$ to be a nested
sequence

\begin{equation}
F_\bullet: 0 = F_0 \subset F_1 \subset F_2 \subset \ldots \subset F_n = E
\end{equation}
with $\dim(F_i) = i$.  For any such flag, we obtain a cell decomposition of $\Gr_k(E)$, as follows.  Let
$\a$ be a partition contained in a $k \times (n-k)$ rectangle (this means that
$\a$ has length at most $k$ and that all parts are less than or equal to $n-k$). To each such $\a$ we
associate the {\em Schubert cell}
\begin{equation}
\O_\a = \{V \in \Gr_k(E)| \dim(V \cap F_j) = i \mbox{ if } n-k+i- \a_i \leq j \leq n-k+i-\a_{i+1} \}.
\end{equation}
and the {\em Schubert variety}
\begin{equation}
X_\a = \{V \in \Gr_k(E)| \dim(V \cap F_{n-k+i-\a_i}) \geq i \}.
\end{equation}
This definition of Schubert cell differs from the one given in the
previous section, but the two definitions refer to the same
object, as we now show.  Given any binary string $\p$ of length
$n$ and weight $k$, associate to it a partition $\a_\p$ as
follows.  Let $a_i$ be the number of zeroes that appear in $\p$
before the $i$th one. Then let $\a_\p = (a_k, a_{k-1}, \ldots,
a_1$).  For instance if $\p = 010011$, then $\a_\p = (3, 3, 1)$.
It is not hard to see that this gives a one-to-one correspondence
between binary strings of length $n$ and weight $k$, and
partitions contained in a $k \times (n-k)$ rectangle, and that
$S_\p = \O_{\a_\p}$.

When we wish to emphasize the flag, we write $\O_\a(F_\bullet)$ and $X_\a(F_\bullet)$ for
$\O_\a$ and $X_\a$, respectively.
Schubert varieties corresponding to partitions with only one nonzero part are called {\em special
Schubert varieties}
\begin{equation}
X_l = \{V \in \Gr_k(E)| V \cap F_{n-k+1-l} \not = 0\}.
\end{equation}

We now show that Schubert varieties are indeed algebraic varieties.  Note that $\dim (V \cap F_i) \geq j$ if and
only if the rank of the map
\begin{equation}
V \hookrightarrow \CC^n \twoheadrightarrow \CC^n/F_i
\end{equation}
is less than or equal to $k-j$.  This means that, in local coordinates,
 all minors of order $k-j+1$ of the matrix of this map must have vanishing determinant,
a requirement governed by polynomial
equations.  The Schubert varieties are therefore algebraic subvarieties of $\Gr_k(E)$.

In what follows, let $f_1, \ldots, f_n$ be a basis respecting the flag $F_\bullet$ of $E$; in other words,
these vectors are such that $F_i = \langle f_1, \ldots f_i \rangle$ for all $i$.
Let $\a$ be a partition contained in a $k \times (n-k)$ rectangle.  In terms of the basis
$\langle f_1, \ldots, f_n \rangle$, any $V \in \O_\a$ can be expressed in terms of a unique basis,
consisting of the rows of a $k \times (n-k)$
matrix with the following properties: the $i$th row contains a $1$ in the
$(n-k+i-\a_i)$th position, and zeros in all subsequent positions; and all other entries in the $(n-k+i-\a_i)$th column
are zero.  For instance, if $n=7, k = 3,$ and $\a = (3, 2, 1)$, such matrices are of the form
\begin{equation}
\left(
\begin{array}{ccccccc}
* & 1 & 0 & 0 & 0 & 0 & 0 \\
* & 0 & * & 1 & 0 & 0 & 0 \\
* & 0 & * & 0 & * & 1 & 0 \\
\end{array}
\right),
\end{equation}
where the stars denote arbitrary entries.  Clearly any such matrix
corresponds to a $V \in \O_\a$, so we have a homeomorphism of
$\O_\a$ with $\CC^{k(n-k)-|\a|}$.  In general, $V$ can be written
(not uniquely) as the span of the rows of any $k \times (n-k)$
matrix with a nonzero entry in the $(n-k+i-\a_i)$th position of
the $i$th row, and zeros afterwards.  Using our example $n=7, k =
3,$ and $\a = (3, 2, 1)$, such matrices can be written as
\begin{equation}
\left(
\begin{array}{ccccccc}
* & * & 0 & 0 & 0 & 0 & 0 \\
* & * & * & * & 0 & 0 & 0 \\
* & * & * & * & * & * & 0 \\
\end{array}
\right),
\end{equation}
where the last star in each row represents any nonzero term, and all other stars represent arbitrary terms.
From this representation, we see that if $\a \subset \b$ (this means that the Young diagram of $\a$ is contained
in the diagram of $\b$), then $\O_\b \subset \overline{\O_\a}$.

The following theorem tells how to determine the incidence of Schubert varieties.
\begin{theorem}
For all partitions $\a \subset k \times (n-k)$,
\begin{itemize}
\item[(a)] $X_\a = \overline{\O_\a} = \coprod_{\b \supset \a} \O_\b$, and
\item[(b)] $X_\b \subset X_\a$ if and only if $\a \subset \b$.
\end{itemize}
\end{theorem}

The Schubert cells $\O_\a$, as a result, form a cellular
decomposition of the Grassmannian.  Therefore, the fundamental
classes of their closures are a basis of the integral cohomology
of $\Gr_k(E)$. (Because all cells are of even real dimension, the
integral cohomology is torsion-free.)  For any Schubert variety
$X_\a$, let $\s_\a = [X_\a]$ denote its class in cohomology,
called a {\em Schubert class}.  The results of this section then
imply the following theorem.


\begin{theorem}
The integral cohomology of the Grassmannian $\Gr_k(E)$ has a basis
given by the Schubert classes $\s_\a$, where $\a$ ranges over all
partitions contained in a $k \times (n-k)$ rectangle:
\begin{equation}
H^*(\Gr_k(E)) = \bigoplus_{\a \subset k \times (n-k)} \ZZ \s_\a.
\end{equation}
The Schubert class $\s_\a$ is an element of $H^{2|\a|}(\Gr_k(E))$.
\end{theorem}



\subsection{Intersections of Schubert Varieties}

Let us now determine when two Schubert varieties must intersect.
Given a flag $F_\bullet$, let $\tilde F_\bullet$ be the opposite flag to $F_\bullet$.  That is,
if $\{f_1, \ldots, f_n\}$ is a basis for $E$ such that $F_k = \langle f_1, \ldots, f_k \rangle$,
then $\tilde F_k = \langle f_{n-k+1}, \ldots, f_n\rangle$.
For any partition $\a$ with at most $k$ rows and $n-k$ columns, let $\O_a = \O_\a(F_\bullet)$ and
let $\tilde \O_\a = \O_\a(\tilde F_\bullet)$.  Because $GL(E)$ acts transitively on the flags,
$\O_\a$ and $\tilde \O_\a$ have the same fundamental class, denoted $\s_\a$.

We have seen that any element of $\O_\a$ can be written as the span of the rows of a unique $k \times (n-k)$
matrix of the form
\begin{equation}
\left(
\begin{array}{cccccccccccccc}
* & \ldots & * & 1 & 0 & \ldots & 0 & 0 & \ldots & \ldots & 0 & \ldots & \ldots & \ldots \\
* & \ldots & * & 0 & * & \ldots & * & 1 & 0 & \ldots & 0 & \ldots & \ldots & \ldots \\
\ldots & \ldots & \ldots & \ldots & \ldots & \ldots & \ldots & \ldots & \ldots & \ldots & \ldots & \ldots & \ldots & \ldots\\
* & \ldots & * & 0 & * & \ldots & * & 0 & * & * & 1 & 0 & \ldots & 0 \\
\end{array}
\right) ,
\end{equation}
where the $i$th row has a $1$ in the $(n-k+i-\a_i)$th position.
Similarly, each element of $\tilde \O_\b$ can be written in terms of a basis whose
elements are the rows of a unique $k \times (n-k)$ matrix of the form
\begin{equation}
\left(
\begin{array}{cccccccccccccc}
0 & \ldots & 0 & 1 & * & * & 0 & * &  \ldots & * & 0 & * & \ldots & * \\
\ldots & \ldots & \ldots & \ldots & \ldots & \ldots & \ldots & \ldots & \ldots & \ldots & \ldots & \ldots & \ldots & \ldots\\
\ldots & \ldots & \ldots & 0 & \ldots & 0 & 1 & * & \ldots & * & 0 & * & \ldots & *\\
\ldots & \ldots & \ldots & 0 & \ldots & \ldots & 0 & 0 & \ldots & 0 & 1 & * & \ldots & *\\
\end{array}
\right) ,
\end{equation}
where the $i$ row has a $1$ in position $\b_{n-k-i+1} + i$.

If $\O_\a \cap \tilde \O_\b \not = \emptyset$, then there must be a $k$-plane $W$ such
that each of the two above matrices
determines a basis for $W$.  Now, the first row of the first matrix cannot be a linear combination of
rows of the second unless $\b_{n-k} + 1 \leq n + 1 - \a_1 \Longrightarrow \a_1 + \b_{n-k} \leq n$.  In general,
in order for the $i$ row of the first matrix to be a linear combination of rows of the second matrix, but not a linear
combination of the first $i-1$ rows of the second matrix, we must have that $\a_i + \b_{n-k-i+1} \leq n$.

For any partition $\a$ contained in an $k \times (n-k)$ rectangle, define $\hat \a$ to be the complementary
partition of $\a$ in the rectangle: that is, $\hat \a_i = n-\a_{n-k-i +1}$.  (If the Young diagram of $\hat \a$ is
turned upside down, it fits perfectly with the diagram of $\a$ to form a $k \times (n-k)$ rectangle.)  The argument
of the previous paragraph shows that $\O_\a \cap \tilde \O_\b = \emptyset$ unless $\b \subset \hat \a$.  We now have

\begin{theorem}
\label{thm:duality}
Suppose $\a$ and $\b$ are two partitions with at most $k$ rows and $n-k$ columns, and that
$|\a| + |\b| = k(n-k)$.  Then the cup product in cohomology of the fundamental classes corresponding to $\a$ and $\b$ is
zero unless $\b = \hat \a$, in which case it is one; that is,
\begin{equation}
\s_\a \cup \s_\b = \d_{\b, \hat \a}.
\end{equation}
The classes $\s_\a$ and $\s_{\hat \a}$ are therefore said to be {\em dual}.
\end{theorem}

{\bf Proof \/ } We have seen that $\O_\a \cap \tilde \O_\b = \emptyset$
unless $\a_i + \b_{n-k+i-1} \leq n$ for all $i$.
Since $|\a| + |\b| = k(n-k)$, we must have equality hold in all these inequalities in order for them
to be simultaneously satisfied, and so $\O_\a \cap \tilde \O_\b = \emptyset$
unless $\b = \hat \a$.  It follows that if $\b \not = \hat \a$, then the intersection of Schubert varieties
$X_\a \cap \tilde X_\b = \emptyset$, so $\s_\a \cup \s_b = 0$.  On the other hand, if $\b = \hat \a$, then
$X_\a \cap \tilde X_\b = \O_\a \cap \tilde \O_\b$.  The above parametrizations of $\O_\a$ and $\tilde \O_\b$
in terms of matrices show that $\O_\a$ intersects $\tilde \O_\b$ in exactly one point, determined by the
basis vectors corresponding to the positions of the $1$'s in both of these matrices.  Now the stars in the
matrices correspond to local coordinates of $\O_\a$ and $\O_\b$; taking all the stars together yields coordinates
for a neighborhood of the intersection in the Grassmannian.  The intersection is obtained at the point where
all coordinates are equal to zero, so it follows that the intersection of $\O_\a$ and $\O_\b$ is transverse at that
point.  Therefore, $\s_\a \cup \s_{\hat a} = 1$.
\hfill
$\Box$

This observation is the starting point for determining the
multiplication rule for Schubert varieties and illustrates the
convenience of dealing with intersections between varieties
associated with flags opposite to each other.
For an integer $l$ between $1$ and $n-k$, let $\s_l$ denote the Schubert class corresponding to the special
Schubert variety $X_l$.  Then the Pieri rule holds for Schubert classes:

\begin{theorem}[Pieri rule for Schubert classes]
\label{thm:pierischubert}
Let $a$ be a partition contained in an $k \times (n-k)$ rectangle, and let $l$ be an integer between
$1$ and $n-k$.  Then
\begin{equation}
\label{eqn:pierischubert}
\s_\a \cup \s_l = \sum_{\n \subset k \times (n-k), \n \in \l \ox k} \s_\n.
\end{equation}
\end{theorem}

The proof even of this theorem consists only of linear algebra,
but is too lengthy to include here. Because the Schubert classes
in cohomology satisfy the Pieri rule, we have the following
result.

\begin{corollary}
The map $\L_k \longrightarrow H^*(\Gr_k(E))$, which sends the Schur function $s_\a$ to the Schubert class
$\s_\a$ if $\a$ is a partition contained in a $k \times (n-k)$ rectangle, and sends $s_\a$ to zero otherwise,
is a surjective ring homomorphism.
\end{corollary}




\section{Computing $\phi^*$} \label{sec:phi}


Using Theorem~\ref{thm:varprin} we can obtain inequalities
relating an operator $\r_{AB}$ and its partial trace $\r_A$
whenever there is a non-empty intersection of the Schubert variety
$X_\b(F)$ with $\phi(X_\a(F'))$, where $F$ and $F'$ are the flags
determined by eigenbases of $\r_{AB}$ and $\r_A$, respectively.
The condition that there must be a nonzero intersection
corresponds cohomologically to there being nonzero product of the
Schubert classes,  $\s_\a \cup \phi^*(\s_\b) \not = 0$, where
$\phi^*: H^*(\Gr_{d_B k}(A \ox B)) \longrightarrow H^*( \Gr_k(A))$
is the map on cohomology induced by $\phi$. In order to compute
when this product is nonzero, we wish to know the behavior of
$\phi^*$, which is easier to determine using another presentation
for the ring $H^*(\Gr(k,n))$, in terms of Chern classes of vector
bundles. In this section we develop this presentation, show how it
corresponds to the previous description of $H^*(\Gr(k,n))$ in
terms of fundamental classes of Schubert varieties, and use it to
describe how $\phi^*$ acts on $H^*(\Gr_{d_B k}(A \ox B))$.


\subsection{Vector Bundles}

Recall that if $M$ is a manifold, then
a $d$-dimensional complex {\em vector bundle} is a map $p:E \rightarrow M$ such
that the fiber $E_p \equiv p^{-1}(b)$ is an $d$-dimensional complex vector space for each $b \in M$, and the following
local triviality condition is satisfied: there is an open cover $\{U_\a\}$ of $M$, together with homeomorphisms
\begin{equation}
h_\a: p^{-1}(U_\a) \rightarrow U_\a \times \CC^d
\end{equation}
that are vector space isomorphisms on each fiber.  Often the total space $E$ is referred to as the vector bundle, with
the rest of the bundle structure implicit.  If $d = 1$, then $E$ is also referred to as a {\em line bundle}.

We will use several standard constructions of bundles:

\begin{itemize}

    \item[(1)] For any manifold $M$, and any $d$, there is the {\em trivial} or {\em product} bundle
    $E = M \times \CC^d$, where $p$ is the projection onto the first factor.

    \item[(2)] If $E$ and $E'$ are bundles, then their direct sum $E \oplus E'$, their tensor product
    $E \ox E'$, and the dual $E^*$ are all defined in a natural way \cite{BT82}.

    \item[(3)] Let $M$ and $N$ be manifolds and $p: E \rightarrow M$ a vector bundle over $M$.  Then
    if $f:N \rightarrow M$ is a (continuous) map, it induces a vector bundle $f^*(E)$ on $N$, given by
    the following subset of $N \times E$:
    \begin{equation}
    \{(n,e):f(n) = p(e)\}.
    \end{equation}
    This bundle  $f^*(E)$, called the {\em pullback of $E$ by $f$},
    is the unique maximal subset of $N \times E$ that makes the following diagram commute:
    \[
       \begin{CD}
       {f^*(E)} @> {} >> {E} \\
       @ V{} VV @ VV {p} V \\
       {N} @>> {f} > {M.} \\
       \end{CD}
       \]

    \item[(4)] Let $V$ be a $d$-dimensional complex vector space and let $\PP(V)$ be its projectivization, that
    is, $\PP(V) = \Gr_1(V)$ is the set of one-dimensional subspaces of $V$.  Let $\hat V$ be the product
    bundle $\PP(V) \times V$.
    Then the {\em universal subbundle}
    $S$ is the subbundle of $V$ given by
    \begin{equation}
    S = \{(\ell, v) \in \PP(V) \times V | v \in \ell \},
    \end{equation}
    also called the {\em tautological line bundle};
    and the {\em universal quotient bundle} $Q$ is defined by the exact sequence
    \begin{equation}
    0 \rightarrow S \rightarrow \hat V \rightarrow Q \rightarrow 0.
    \end{equation}
    This is known as the {\em tautological exact sequence} over $\PP(V)$.  The dual $S^*$ is called the
    {\em hyperplane bundle}.

\end{itemize}

We will also use the following fact \cite{Hatcher}.
\begin{proposition}
\label{thm:split}
Let $0 \rightarrow A \rightarrow B \rightarrow C \rightarrow 0$ be an exact sequence of
vector bundles.  Then $B$ is isomorphic as a bundle to $A \oplus C$.
\end{proposition}

Instead of requiring the fiber of each point of a manifold $M$ to be a vector space in our definition,
 we may have it be any topological
space $F$, thus obtaining a {\em fiber bundle} with fiber $F$
\cite{Hatcher}. The main example of this will be the {\em
projective bundle}  $\PP(E) \rightarrow B$ associated to any
$d$-dimensional vector bundle $E \rightarrow B$.  The fiber at
each point of $\PP(E)$ is isomorphic to the complex projective
space $\PP^{d-1}$, and the local trivializations of $\PP(E)$ are
induced by those of $E$ \cite{BT82}.  If we let $p$ denote the
projection from $\PP(E)$ to $M$, then we may pull back $E$ by $p$
to obtain a bundle $p^*(E)$ over $\PP(E)$, whose fiber at any
point $\ell_p$ is $E_p$. As in example~(4) above, this pullback
bundle has a universal subbundle
$S = \{(\ell_p, v) \in p^*(E)|v \in \ell_p\} $
and a universal quotient bundle $Q$ defined by exactness of the sequence
$0 \rightarrow S \rightarrow p^*(E) \rightarrow Q \rightarrow 0$.

\subsection{Chern Classes}

{\em Chern classes} are integral cohomology classes naturally
associated to complex vector bundles. We will need the following
fact. Let $\PP^d$ be the $d$-dimensional complex projective space.
Since $\PGL_{d+1}$ is a connected group acting transitively on the
hyperplanes of $\PP^d$, the fundamental class in cohomology
associated to a hyperplane $H$ does not depend on the chosen
hyperplane.  Let $h$ denote this class, which we call the {\em
hyperplane class}.

Chern classes are defined axiomatically as follows~\cite{Hatcher}:

\begin{theorem}
There are unique functions $c_1, c_2, \ldots$ on complex vector
bundles $E \rightarrow N$ over compact differentiable varieties,
with $c_i(E) \in H^{2i}(M)$, that depend only on the isomorphism
type of $E$ and satisfy the following properties:
\begin{itemize}
    \item[(a)] (functoriality) For any continuous map $f:N \rightarrow M$,
            $c_i(f^*(E)) = f^*(c_i(E))$.
    \item[(b)] (Whitney sum formula) Writing $c = 1 + c_1 + c_2 + \ldots$, we have
            $c(E_1 \oplus E_2) = c(E_1) \cup c(E_2)$.
    \item[(c)] If $i > \dim E$, then $c_i(E) = 0$.
    \item[(d)] (normalization) For the tautological line bundle $S$ on $\PP^d$, $c_1(S) = -h$, the
        negative of the hyperplane class.
\end{itemize}
These classes $c_i(E)$ are called {\em Chern classes} of the vector bundle $E$, and
$c(E) = \sum_k c_k(E)$ is called the {\em total Chern class} of $E$ (setting $c_0(E) = 1$).
\end{theorem}

We note that the Whitney sum formula may be written as
\begin{equation}
c_k(E \oplus F) = \sum_{i+j=k} c_i(E) \cup c_j(F).
\end{equation}
It can be shown \cite{Hatcher} that the axiomatic properties of Chern classes imply that if $L_1$ and $L_2$
are line bundles, then $c_1(L_1 \ox L_2) = c_1(L_1) + c_1(L_2)$.  From this fact, it readily follows that
$c_1(L) = 0$ if $L$ is a trivial line bundle, and hence that $c_k(E)$ is zero for any trivial bundle $E$, by
the Whitney formula.

We now specialize to the problem at hand. Let $T$ be the
tautological bundle of dimension $k$ over $\Gr(k,n)$, for which
the fiber over a subspace $V$ is $V$ itself.  Let $Q$ be the
quotient bundle over $\Gr(k,n)$ whose fiber over a vector space
$V$ is $\CC^n/V$.  Then the properties of Chern classes imply the
following result \cite{Manivel01}.

\begin{theorem}
\label{thm:chernq}
The $l$th Chern class of the quotient bundle, $c_l(Q)$, is equal to the class of the special Schubert variety
$\s_l$.
\end{theorem}
{\bf Proof \/ }
Fix a complete flag $F_\bullet$ for the $n$-dimensional complex vector space.
Let $\a(1,l)$ be the partition corresponding to the complement of a $1 \times l$ rectangle
in the $k \times (n-k)$ rectangle.  We must show that for any partition $\a \subset k \times (n-k)$
of weight $k(n-k) - l$,
$c_l(Q) \cup \s_\a = 1$ if $\a = \a(1, l)$, and $c_l(Q) \cup \s_\a = 0$ otherwise.

Suppose that $\a$ has weight $k(n-k) - l$ but $\a \not = \a(1,l)$.  Then $\a_k \geq n-k -l + 1$, so any
$V \in X_\a$ satisfies $\dim(V \cap F_{k+l-1}) \geq k$.  This means that $V \subset F_{k+l-1}$, so that
$X_\a$ is contained in the smaller Grassmannian $G = \Gr(k, k+ l-1)$
of $k$-dimensional subspaces of $F_{k+l-1}$.  Let $j: G \hookrightarrow G(k, n)$ be the inclusion map.
Using the projection formula from topology \cite{Manivel01}, we have that
\begin{equation}
\label{eq:proj}
c_l(Q) \cup \s_\a = j_*(j^*(c_l(Q)) \cup [X_\a]),
\end{equation}
where $j_*$ is the Gysin homomorphism on cohomology arising from Poincar\'e duality.  But by the
exact sequence of bundles over $G$,
\begin{equation}
0 \rightarrow F_{k+l-1}/V \rightarrow \CC^n/V \rightarrow \CC^n/F_{k+l-1} \rightarrow 0,
\end{equation}
the restriction $Q_G$
of the quotient bundle to $G$ can be written $Q_G = F_{k+l-1}/V \oplus \CC^n/F_{k+l-1}$,
where the latter bundle in the direct sum is trivial.  It follows from the Whitney formula that
$c_l(Q_G) = 0$, so since $c_l(Q_G) = j^*(c_l(Q))$, we must have that
$c_l(Q) \cup \s_\a = 0$ by Equation~\ref{eq:proj}.

Now suppose that $\a = \a(1,l)$.  In this case
\begin{equation}
X_\a = \{V \in \Gr(n,k)| F_{k-1} \subset V \subset F_{k-l}\},
\end{equation}
which is isomorphic to the $l$-dimensional projective space $\PP = \PP(F_{k+l}/F_{k-1})$.
Let $i$ denote the natural isomorphism from $X_\a$ to $\PP$.  On $\PP$ we have the
exact sequence
\begin{equation}
0 \rightarrow V/F_{k-1} \rightarrow F_{k+l}/F_{k-1} \rightarrow Q_\PP \rightarrow 0.
\end{equation}
Here $ V/F_{k-1}$ is the tautological line bundle, $Q_\PP$ is the quotient bundle,
and $ F_{k+l}/F_{k-1}$ is a trivial bundle.  It follows that the total Chern class
of $Q_\PP$ is $c(Q_\PP) = (1-h)^{-1}$ (where $h$ is the class of the hyperplane).
Now the projection formula tells us that
\begin{eqnarray*}
c_l(Q) \cup \s_\a & = & i_*(i^*(c_l(Q))\cup [X_\a]) \\
& = & i_*(i^*(c_l(Q)) \\
& = & i_*(c_l(Q_\PP)) \\
& = & 1.
\end{eqnarray*}
\hfill
$\Box$

\subsection{The Splitting Principle}

We have seen that the Chern classes of the quotient bundle $Q$
correspond to special Schubert classes. Since all Schubert classes
can be obtained as products of these special Schubert classes,
characterizing the effect of $\phi^*$ on the Chern classes of $Q$
will be sufficient to determine the effect of $\phi^*$ on
$H^*(\Gr(k,n)).$  To do this, we will need the {\em splitting
principle}, a fundamental observation from the theory of Chern
classes. In what follows, let $E$ be any vector bundle over a
manifold $M$, whose dimension we denote by $m$. We shall have in
mind the case where $M = \Gr(k,n)$ and $E$ is the quotient bundle
$Q$ defined above (so that $m = n-k$).

Starting with the bundle $E$ over $M$, let $\PP(E)$ be the
projectivization of $E$, and let $f_1$ be the induced map from
$\PP(E)$ to $M$.  Let $f_1^*(E)$ be the pullback bundle:

    \[
       \begin{CD}
       {f_1^*(E)} @> {} >> {E} \\
       @ V{} VV @ VV {} V \\
       {\PP(E)} @>> {f_1} > {M.} \\
       \end{CD}
       \]
Let $L_1$ be the tautological line bundle of the pullback $f^*(E)$.
Then we have an exact sequence
\begin{equation}
0 \rightarrow L_1 \rightarrow f_1^*(E) \rightarrow Q_1 \rightarrow 0,
\end{equation}
where $E$ is an $(m-1)$-dimensional bundle over $M$, so $f_1^*(E)$
is isomorphic to $L_1 \oplus Q_1$. Similarly, let $\PP(Q_1)$ be
the projectivization of $Q_1$, with $f_2$ as the map from
$\PP(Q_1)$ to $\PP(Q)$.  If $L_2$ is the tautological line bundle
of $\PP(Q_1)$, then $L_2$ gives rise to a quotient $Q_2$ such that
$f_2^*(Q_1)$ is isomorphic to $L_2 \oplus Q_2$.  We can thus pull
back $E$ to a direct sum of $Q_2$ and two line bundles:

\begin{equation}
\label{eq:split}
\xymatrix{
 & & f_2^*(L_1) \oplus L_2 \oplus Q_2 \ar[d]\\
 & L_1 \oplus Q_1 \ar[d]& \PP(Q_1) \ar[ld]_{f_2}\\
E \ar[d] & \PP(E) \ar[ld]_{f_1}& \\
M & & }
\end{equation}

Continuing in this way, we obtain bundles $Q_3, \ldots, Q_{m-1}$,
and projectivizations $\PP(Q_2), \ldots,$ $\PP(Q_{m-2})$, such
that the pullback of $E$ by the map from $\PP(Q_{m-2})$ to $M$ is
a direct sum of line bundles.  If $f = f_1 \circ f_2 \circ \ldots
f_{m-2}$ is the map from $\PP(Q_{m-2})$ to $M$, then it can be
shown that the induced map on cohomology $f^*:H^*(M) \rightarrow
H^*(\PP(Q_{m-2}))$ is injective \cite{BT82}.  We summarize these
facts in the following theorem, known as the splitting principle:.

\begin{theorem}[The Splitting Principle]
\label{thm:splitprin}
For any vector bundle $E$ on a manifold $M$, there exists a manifold $N$ and a continuous
$f:N \rightarrow M$ such that $f^*(M) \rightarrow f^*(N)$ is injective, and pullback bundle $f^*(E)$
is a direct sum of line bundles.
\end{theorem}



We now illustrate the splitting principle by using it to derive a result that will be useful to us.
Let $E$ be a vector bundle, and let $f: N \rightarrow M$ be the map given by Theorem~\ref{thm:split},
so that the pullback $f^*(E)$ splits as the direct sum of line bundles $L_1, \ldots, L_n$.
Let $x_i = c_1(L_i)$.
Then
the Whitney sum formula $c_k(E_1 \oplus E_2) = \sum_{i+j=k} c_i(E_1) \cup c_j(E_2)$ implies that
\begin{equation}
c_k(f^*(E)) = c_k(x_1, \ldots, x_n)
\end{equation}
is the $k$th elementary symmetric polynomial in the first Chern classes of $f^*(E)$.  By the
functoriality of the Chern classes, it follows that $f^*(c_k(E))$ is the $k$th elementary symmetric
polynomial in $c_1(L_1), \ldots, c_1(L_n)$.

Let us revisit the construction of the split manifold of a vector
bundle $E$.  $\PP(E)$ consists of pairs $(x, \ell)$, where $x \in
M$ and $\ell$ is a line in $E_x$.  Proposition~\ref{thm:split}
allows us to consider all the bundles $Q_1, \ldots Q_{n-1}$ as
subbundles of $E$.  Now $\PP(Q_1)$ consists of triples $(x,
\ell_1, \ell_2)$ where $\ell_2$ is a line in the linear complement
of $\ell_1$ in $E_p$.  In general, a point of $\PP(Q_j)$ over $(x,
\ell_1, \ldots, \ell_j)$ in $\PP(Q_{j-1})$ is a $(j+2)$-tuple $(x,
\ell_1, \ldots, \ell_j, \ell_{j+1})$ where $\ell_{j+1}$ is a line
in the complement of $\ell_1, \ldots, \ell_j$.  We conclude that
the split manifold $\PP(Q_{m-2})$ is in fact the {\em flag
bundle}:
\begin{equation}
\Fl(E) = \{(x, \ell_1 \subset \langle \ell_1, \ell_2 \rangle \subset \langle \ell_1, \ell_2, \ell_3 \rangle \subset \ldots
\subset E_x)| x \in M\}.
\end{equation}

\subsection{Representations and Line Bundles}

We have seen that the splitting principle allows us to regard the
Chern classes of a vector bundle $E$ as (symmetric) polynomials in
the first Chern classes of the line bundles of a flag bundle
associated to $E$. Given an $m$-dimensional vector space $V$, the
space $\Fl (V)$ of all complete flags on $V$ can be identified
with $\GL(V)/T$, where $T$ is now the group of upper triangular
matrices. This follows because $\GL(V)$ is transitive on the flags
and $T$, the stabilizer of the standard flag $0 \subset \langle
e_1 \rangle \subset \langle e_1, e_2 \rangle \subset \cdots
\subset \langle e_1, \ldots, e_m \rangle = V$, is isomorphic to
the stabilizer of any given flag. We can associate to any
one-dimensional representation $\c:T \rightarrow \CC^*$ a line
bundle over the flag manifold $\Fl (V)$ as follows:
\begin{equation}
L(\c) = \GL(V) \times \CC/((g t, z) \sim (g, \c(t)z))
\end{equation}
for $g \in \GL(V)$, $t \in T$, and $z \in \CC$. The projection of
$L(\c)$ onto $\Fl (V)$ is just $(g, z) \buildrel \pi \over
\rightarrow (gT)$. Under the action of $\GL(V)$ given by $h(gt, z)
= (hgt, z)$, the following diagram commutes:

\[
       \begin{CD}
       {L(\c)} @> {GL(V)} >> {L(\c)} \\
       @ V{\p} VV @ VV {\p} V \\
       {\Fl (V)} @> {GL(V)} >> {\Fl (V)}, \\
       \end{CD}
       \]
since $(hgt, z) = (hg, \c(t)z)$. The line bundle $L(\c)$ is thus
\emph{equivariant} with respect to the bundle projection.

Conversely, suppose $L$ is an equivariant line bundle over $\Fl
(V)$.  Then $T$ acts on the fiber over $eT$, so this fiber is a
one-dimensional representation $\c$ of $T$. Indeed, the the line
bundle $L(\c)$ corresponding to this representation is isomorphic
to $L$. Let $y \in L$ lie in the fiber over $eT$. The isomorphism
is given by the map $r: L(\c) \rightarrow L$ given by $r(g,z)
= z(g \cdot y)$.  At first glance, the map isn't obviously well-defined,
but note that $r(gt,z) = z(gt \cdot y) = z(g \c(t)y) = r(g,
\c(t)z)$. Since $G$ acts transitively on the fibers of $L$, and
multiplication by $z$ is a surjective map on any given fiber, $r$
is surjective. For injectivity, suppose that $z_1(g_1 \cdot y)
= z_2 (g_2 \cdot y)$.  If $z_1 \not
= 0$, then $y = z_1^{-1}z_2g_1^{-1}g_2 \cdot y$, so $g_1^{-1} g_2
\in T$. This means that $g_2 = g_1 t $ for some $t \in T$ and
$z_1^{-1}z_2 \c(t) = 1$, so $z_2 = z_1\c(t)^{-1}$.  Thus, as
elements of $L(\c)$, $(g_2, z_2) = (g_1 t, \c(t)^{-1}z_1) = (g_1,
z_1)$, so $r$ is indeed injective. It is likewise easy to see that
$r$ commutes with the bundle projection and is linear on fibers.
The correspondence between line bundles and one-dimensional
representations of $T$ is therefore a bijection.

Actually, we can identify the characters $\c$ with line bundles on
a flag manifold $\Fl(V)$ more explicitly. Consider the {\em
tautological filtration} \cite{Fulton97}
\begin{equation}
0 = U_0 \subset U_1 \subset U_2 \subset \cdots \subset U_m = \Fl
(V) \times V
\end{equation}
of vector bundles over $\Fl(V)$, where $\Fl(V) \times V$ is the
product bundle, and $U_k$ is the $k$-dimensional bundle over
$\Fl(V)$ whose fiber over a flag $V_1 \subset \cdots \subset V_m$
is $V_k$. It follows from the splitting principle that the
cohomology ring $H^*(\Fl(V))$ is generated by the first Chern
classes of the line bundles $L_i = U_i/U_{i-1}$, setting $x_i =
c_1(L_1)$.  The identity matrix fixes the standard flag $\{e_1,
\ldots, e_m\}$.  Therefore, over $eT$, the fiber of $L_i$ is
$V_i/V_{i-1}$, where $V_i = \langle e_1, \ldots, e_i \rangle$. If
$v = \sum_{k=1}^i \a_i e_i \in V_i$ and $t \in T$, then $t \cdot v
= w + t_{ii} e_i$, where $w \in V_{i-1}$ and $t_{ii}$ is the $i$th
diagonal entry of $t$.  We have shown the following:
\begin{theorem}
If $L_i$ is the line bundle over a flag manifold defined as above,
then the character $\c$ associated to $L_i$ is the map taking $t$
to $t_{ii}$.
\end{theorem}

Let us adapt this machinery to the problem at hand.  Recall that
we have two complex vector spaces $A$ and $B$ of dimensions $d_A$
and $d_B$, respectively, together with a map $\phi:\Gr_k(A)
\rightarrow \Gr_{kd_B}(A \ox B)$ given by $\phi(V) = V \ox B$. We
wish to compute the action of the induced map $\phi^*:
H^*(\Gr_{kd_B}(A \ox B)) \rightarrow H^*(\Gr_k(A))$.

Let $Q_A$ and $Q_{AB}$ be the quotient bundles of
Theorem~\ref{thm:chernq} over the Grassmannians $\Gr_k(A)$ and
$\Gr_{kd_B}(A \ox B)$ respectively.  The Chern classes of these
bundles are the classes of the special Schubert varieties in the
cohomology rings.  By the splitting principle, the associated flag
bundles $\Fl(A)$ and $\Fl(A \ox B)$ have pullbacks which split as
a direct sum of line bundles $L_i$ of the respective tautological
filtrations.  The cohomology of the Grassmannians embeds in the
cohomology of these pullbacks, so we may determine $\phi^*$ by its
action on the Chern classes of the pullback bundle of  $\Fl(A \ox
B)$.

 It follows from the definition of pullback bundles that the bundle
$\phi^*(L(\c_i))$ is the set of triples $(gT_A, \phi(g), z) \in
\GL(A)/T_A \times \GL(A \ox B) \times \CC$ with the identification
$(gT_A, \phi(g \cdot t), z) \sim (gT_A, \phi(g), \c(\phi(t)) z)$.
This means that $\phi^*(L(\c_i)) = L(\phi^*(\c_i))$. The pullback
of the map induced by $\phi$ on the characters of the group
$T_{AB}$ is readily computed: for a matrix $X \in T_A$, and the
character $\c_i$ taking a matrix to its $i$th diagonal entry, we
have $\phi^*(\c_i)(X) = \c_i(\phi(X)) =  \c_i(X \ox I) =
\c_{\lceil i/d_B \rceil}(X)$.  So $\phi^*(\c_i) = \c_{\lceil i/d_B
\rceil}$.  Now we can calculate the action of $\phi^*$ on the
Chern classes:
\begin{eqnarray}
\phi^*(x_i)
& = & \phi^*(c_1(L(\c_i))) \nonumber \\
& = & c_1(\phi^*(L(\c_i))) \nonumber \\
& = & c_1(L(\phi^*(\c_i))) \nonumber \\
& = & c_1(L(\c_{\lceil i / d_B \rceil})).
\end{eqnarray}
Since that short calculation was the reason for developing so much
machinery, we give the conclusion the status of a theorem:
\begin{theorem}
$\phi^*(x_i) =  c_1(L(\c_{\lceil i / d_B \rceil}))$.
\end{theorem}

\section{Determining the Inequalities} \label{sec:inequalities}

In this section we use our knowledge of how $\phi^*$ behaves to
explicitly derive inequalities relating the spectra of $\r_{AB}$
and of $\r_A$ and work out some examples in low dimensions. We
also restate how to obtain the inequalities in the language of
representation theory.  Next, we discuss recent progress in
symplectic geometry that shows that the inequalities derived using
the method described here are sufficient. Finally, we prove that
if $d_B \geq \frac{1}{2}d_A^2$, then the inequalities simplify
greatly.

\subsection{Putting It All Together}
\label{sec:together}

Let $\r_A = \Tr_B \r_{AB}$, and let $\l$, $\m$, and $\tilde \l$ denote the spectra of
$\r_{AB}$, $-\r_{AB}$, and $\r_A$, respectively.
Theorem \ref{thm:varprin} can be interpreted cohomologically as saying that if

\begin{equation}
\phi^*(\s_\p) \cup \tilde \s_\n \not = 0,
\end{equation}
where $\s_\p \in H^*(\Gr(kd_B, d_Ad_B))$ and $\tilde \s_\n \in H^*(\Gr(k, d_a))$ are
Schubert classes,
then the spectra $\m$  and $\tilde \l$
must satisfy the inequalities
\begin{equation}
\label{eqn:eigineq}
\sum \n(i) \tilde \l_i + \sum \p(i) \m_i \leq 0.
\end{equation}
Now $\phi^*(\s_\p)$ is an integer combination of Schubert classes,
\begin{equation}
\phi^*(\s_\p) = \sum_i n_i \tilde \s_{\p_i}.
\end{equation}
For each of these classes, $\tilde \s_{\p_i} \cup \tilde \s_\n
\not = 0$ iff $\n$ contains the complement of $\p_i$ in the $k
\times (n-k)$ rectangle.  But if we consider the case where $\n$
is in fact the complement of $\p_i$, then we see that the
Inequalities~(\ref{eqn:eigineq}) are the strongest in this case;
for any other $\n' \supset \n$, the inequalities determined by
$\n'$ are implied by the inequalities determined by $\n$. So it is
sufficient to consider complements of each Schubert class $\tilde
\s_{\p_i}$ contained in $\phi^*(\s_\p)$, in order to obtain the
inequalities relating $-\r_{AB}$ and $\r_A$. Now if $\m$ is the
spectrum of $-\r_{AB}$, then the spectrum $\l$ of $\r_{AB}$ is
given by $\l_i = -\m_{d_A-i+1}$ (since the ordering of the
eigenvalues is reversed).  Given binary strings $\p, \hat \p \in
{d_Ad_B \choose k}$ satisfying $\hat \p(i) = \p(d_Ad_B -i + 1)$,
so that $\hat \p$ is simply the string $\p$ in reverse, the
Schubert cell $S_{\hat \p}$ corresponds to the complementary
partition to that of $S_\p$.  This means that we obtain
inequalities
\begin{equation}
\label{eqn:goodineq}
\sum \n(i) \tilde \l_i \leq \sum \p(i) \l_i
\end{equation}
whenever $\phi^*(\s_{\hat \pi})$ contains $\s_{\hat \n}$ (where $\hat \n$ is the complementary partition to
$\n$) as a summand.  It then follows
 that Inequalities~(\ref{eqn:goodineq}) are obtained whenever $\phi^*(\s_\p)$ contains
$\s_\n$ as a summand.

Theorem~\ref{thm:chernq} says
that the $l$th Chern class $c_l(Q)$ of the universal quotient bundle $Q$ over
the Grassmannian $\Gr(k,n)$ is equal to the special Schubert class $\s_l \in H^*(\Gr(k,n))$.
And the splitting principle allows us to conclude that
\begin{equation}
c_l(Q) = e_l(x_1, \ldots, x_{n-k}),
\end{equation}
where $x_i = c_1(L_i)$ is the first Chern class of the $i$th split component of $f^*(Q)$, and
$e_l$ is the $l$th elementary symmetric polynomial.  Because the special Schubert classes
$\s_l$ generate the cohomology ring, we therefore have a surjective ring homomorphism
\begin{eqnarray*}
\tilde \psi: & \L_{n-k} & \rightarrow H^*(\Gr(k,n))\\
    & e_l(x_1, \ldots x_{n-k}) & \mapsto \s_l.
\end{eqnarray*}
We may compose the map $\tilde \psi$ with the involution $\o: \L_{n-k} \rightarrow \L_{n-k}$,
$\o(e_k) = h_k$, to obtain a map
\begin{eqnarray*}
\psi: & \L_{n-k} & \rightarrow H^*(\Gr(k,n))\\
    & h_l(x_1, \ldots x_{n-k}) & \mapsto \s_l.
\end{eqnarray*}
Now, by the Pieri rule, it follows that for any partition $\l$,
$\psi(s_\l(x_1, \ldots, x_{n-k})) = \s_\l$.
Thus, we may determine how $\phi^*$ acts on $H^*(\Gr(kd_B, d_A d_B))$ by determining how the
map $x_i \mapsto x_{\lceil i / d_B \rceil}$ acts on Schur functions.

\subsection{Some Observations}
\label{sec:obs}

In this section we make some observations about the map $\phi^*$
that will simplify our computations to some degree.
First, we note that $\phi^*$
is particularly easy to calculate on the Newton power sums $p_j = \sum_i x_i^j$:
\begin{eqnarray}
\phi^*(p_j(x_1, \ldots, x_{(d_A - k) d_B}))& = & \phi^*\Big(\sum_{i=1}^{(d_A-k)d_B} x_i^j\Big)\\
& = & \sum_{i=1}^{(d_A-k)d_B} x_{\lceil i/d_B \rceil}^j\\
& = & \sum_{i=1}^{d_A -k} d_B x_i^j\\
& = & d_B p_j(x_1, \ldots, x_{d_A -k}).
\end{eqnarray}

We further note that the total degree of a polynomial in the Chern
classes $x_1, \ldots x_{n-k}$ is equal to the weight of the
corresponding partition, and $\phi^*$ maps every monomial in $x_1,
\ldots,$ $x_{(d_A -k)d_B}$ to a monomial in $x_1, \ldots,$ $x_{d_A
-k}$ of the same total degree, so that $\phi^*(\s_\p)$ is a sum of
Schubert classes of the same weight as $\pi$.

Applying this observation to the empty partition $\a = (0)$, which corresponds to the binary string
$\underbrace{11 \ldots 1}_k \underbrace{00\ldots 0}_{n-k}$ in $\Gr(k, n)$, we obtain the inequalities
\begin{equation}
\label{eqn:basicineq}
\sum_{i=1}^k \tilde \l_i \leq \sum_{i=1}^{d_B k} \l_i
\end{equation}
for every $k \in \{1, \ldots, d_A\}$. These are the same
inequalities previously derived in Lemma~\ref{lem:basicineq} using
only Ky Fan's Maximum Principle. We will call
Inequalities~(\ref{eqn:basicineq}) {\em basic inequalities}.  As
we shall see, many of the inequalities that arise from considering
the intersections of Schubert classes will not contain additional
information; rather, they will be consequences of the basic
inequalities.  We call such inequalities {\em redundant
inequalities}.

Finally, we argue that it is sufficient to consider inequalities derived from $\phi^*$ acting on
$H^*(\Gr(kd_B, d_A d_B))$,
where $k \leq \frac{d_A}{2}$. To see this, suppose there is an inequality of the form
\begin{equation}
\sum_{i=1}^{d_A} \n(i) \tilde \l_i \leq \sum_{i=1}^{d_A d_B} \p(i) \l_i,
\end{equation}
where the weight of $\n$ is greater than
$\frac{d_A}{2}$.  We may apply this inequality to the matrices $-\r_{AB}$ and $-\r_A$
and use the trace condition to conclude that
\begin{equation}
\sum_{i=1}^{d_A} \n'(i) \tilde \l_i \leq \sum_{i=1}^{d_Ad_B} \p'(i) \l_i,
\end{equation}
where $\n'(i) = 1 - \n(i)$ for all $i$, and similarly for $\p'$.  If the weight of
$\n$ is greater than $\frac{d_A}{2}$, then the weight of $\n'$ is less than
$\frac{d_A}{2}$.  Thus, the desired inequality is a consequence of an inequality involving
fewer than $\frac{d_A}{2}$ eigenvalues.  (This argument is not valid unless we know
that our method generates all possible valid inequalities.  This is indeed the case, but
we postpone the discussion for Section~\ref{sec:sufficiency}.)

\subsection{Examples}

We now work out the inequalities for some examples.  The case $d_A = 2$ was already solved in
Section~\ref{sec:dim2sol}, where it was shown that the basic inequalities were the only
constraints on the eigenvalues of $\r_A$ and $\r_{AB}$.  Thus, the simplest remaining case
is $d_A = 3$, $d_B = 2$, which we will now illustrate.  We use $h_l$ to refer to the
$l$th complete symmetric function, and $p_l$ to refer to the $l$th Newton power sum
symmetric function.  We identify Schur functions with their images as Schubert classes, denoting
 either by a (Young diagram of a) partition.

As we have argued, we may restrict attention to inequalities involving at most $\frac{d_A}{2}$
eigenvalues; in the case $d_A = 3$, this means that it suffices to consider maps
$\phi^*: H^*(\Gr(2, 6)) \rightarrow H^*(\Gr(1, 3))$.  The Schubert classes of $H^*(\Gr(1, 3))$
correspond to partitions that fit inside a $1 \times 2$ rectangle, of which there are only
two (excluding the empty partition, for which we obtain the basic inequalities):
{\tiny $\yng(1)$} and {\tiny $\yng(2)$}.  Because $\phi^*$ preserves the weight of a partition, we
need only consider partitions of weight one and two in  $H^*(\Gr(2, 6))$: namely,
{\tiny $\yng(1)$}, {\tiny $\yng(2)$}, and {\tiny $\yng(1,1)$}. Figure~\ref{fig:3by2} lists
 the Schur polynomials and binary strings associated to each of these partitions (the polynomials
are readily computed using the Jacobi-Trudi formula).
\begin{figure}
\begin{center}
\begin{tabular}{|c|c|c|c|}
\hline
$\a$ & $s_\a$ & $\p_\a \in H^*(\Gr(2, 6))$ & $\p_\a \in H^*(\Gr(1, 3))$ \\
\hline
{\tiny $\yng(1)$} & $p_1$ & $101000$ & $010$\\
{\tiny $\yng(2)$} & $\frac{1}{2}(p_1^2 + p_2)$ & $100100$ & $001$\\
{\tiny $\yng(1,1)$} & $\frac{1}{2}(p_1^2 - p_2)$ & $011000$ & --- \\
\hline
\end{tabular}
\caption{\label{fig:3by2}
Partitions, their Schur polynomials and binary strings}
\end{center}
\end{figure}

Using this information, we can calculate $\phi^*$ on each of the Schubert classes
 {\tiny $\yng(1)$}, {\tiny $\yng(2)$}, and {\tiny $\yng(1,1)$} $\in H^*(\Gr(2, 6))$
:

\begin{itemize}

\item[(1)] $\phi^*({\tiny \yng(1)}) = \phi^*(p_1) = 2 p_1 = 2 {\tiny \yng(1)}$.  This
yields the inequality $\tilde \l_2 \leq \l_1 + \l_3$.

\item[(2)] $\phi^*({\tiny \yng(2)}) = \phi^* (\frac{1}{2}(p_1^2 + p_2)) = 2 p_1^2 + p_2 =
3 {\tiny \yng(2)} + {\tiny \yng(1,1)}$.  For the {\tiny $\yng(2)$} term on the right side,
we get the inequality $\tilde \l_3 \leq \l_1 + \l_4$.  The  {\tiny $\yng(1,1)$} term does not
yield an inequality because  ${\tiny \yng(1,1)} = 0$ in  $H^*(\Gr(1, 3))$.

\item[(3)] $\phi^*( {\tiny \yng(1,1)}) = \phi^* (\frac{1}{2}(p_1^2 - p_2)) = 2 p_1^2 - p_2 =
3 {\tiny \yng(1,1)} + {\tiny \yng(2)}$.  As before, the {\tiny $\yng(1,1)$} term does not yield an
inequality.  The {\tiny $\yng(2)$} term yields the inequality $\tilde \l_3 \leq \l_2 + \l_3$.

\end{itemize}

So we have three inequalities,  $\tilde \l_2 \leq \l_1 + \l_3$,
$\tilde \l_3 \leq \l_1 + \l_4$, and $\tilde \l_3 \leq \l_2 +
\l_3$.  Let us check these inequalities for redundancy. From the
basic inequalities, we have that $\tilde \l_2 \leq \frac{1}{2}
(\tilde \l_1 + \tilde \l_2) \leq \frac{1}{2} (\l_1  + \l_2 + \l_3
+ \l_4) \leq \l_1 + \l_3$, so the first inequality is redundant.
And $\tilde \l_3 \leq \frac{1}{3} (\tilde \l_1 + \tilde \l_2 +
\tilde \l_3) \leq \frac{1}{3} (\l_1 + \l_2 + \l_3 + \l_4 + \l_5 +
\l_6) \leq \l_1 + \l_4$, so the second inequality is also
redundant.  However, the inequality $\tilde \l_3 \leq \l_2 + \l_3$
is not redundant. (For example, $\l = (1, 0, 0, 0, 0, 0)$ and
$\tilde \l = (\frac{1}{3}, \frac{1}{3}, \frac{1}{3})$ satisfy the
basic inequalities, but $\tilde \l_3 \leq \l_2 + \l_3$ does not.)

So  $\tilde \l_3 \leq \l_2 + \l_3$ is the only new inequality we get involving one eigenvalue of
$\r_A$.  By duality, we also have the inequality $\tilde \l_2 + \tilde \l_3 \leq \l_1 + \l_2 + \l_3 +
\l_6$, or
$\tilde \l_1  \geq \l_4 + \l_5$.  Thus, our complete list of eigenvalue constraints on $\r_{AB}$
and $\r_A$ is
\begin{eqnarray}
\tilde \l_1 & \leq & \l_1 + \l_2, \\
\tilde \l_3 & \geq & \l_5 + \l_6, \\
\tilde \l_3 & \leq & \l_2 + \l_3,\\
\tilde \l_1 & \geq & \l_4 + \l_5,
\end{eqnarray}
together with the trace condition $(\tilde \l_1 + \tilde \l_2 + \tilde \l_3) =
(\l_1 + \l_2 + \l_3 + \l_4 + \l_5 + \l_6)$.

Now we consider the case $d_A = 3$, $d_B = 3$.  We have that
\begin{eqnarray}
\phi^*({\tiny \yng(1)})&  = & 3 {\tiny \yng(1)}, \\
\phi^*({\tiny \yng(2)})&  = & 6 {\tiny \yng(2)} + 3 {\tiny \yng(1,1)}, \\
\phi^*({\tiny \yng(1,1)})&  = & 6 {\tiny \yng(1,1)}+ 3 {\tiny \yng(2)},
\end{eqnarray}
yielding inequalities

\begin{eqnarray}
\tilde \l_2 & \leq & \l_1 + \l_2 + \l_4, \\
\tilde \l_3 & \leq & \l_1 + \l_2  + \l_5, \\
\tilde \l_3 & \leq & \l_1 + \l_3 + \l_4.
\end{eqnarray}
It is not hard to check that all of these inequalities are redundant.  Thus, our only inequalities
for the case $d_A = 3$, $d_b = 3$ are the basic inequalities

\begin{eqnarray}
\tilde \l_1 & \leq & \l_1 + \l_2 + \l_3, \\
\tilde \l_3 & \geq & \l_7 + \l_8 + \l_9.
\end{eqnarray}

\subsection{Representation Theory Perspective}
\label{subsec:repnTheory}

Given a Schur polynomial $s_\l$, we have seen how to determine $\phi^*(s_\l)$ as follows:
write $s_\l$ in terms of Newton power sums, evaluate $\phi^*$ on each of the power sums, and
then express the results in terms of Schur polynomials.  While this algorithm is fairly
straightforward, the relationship between $s_\l$ and the terms appearing
in $\phi^*(s_\l)$ is less clear.  In this section, we see that we can interpret this
relationship from the standpoint of group representation theory.  Asking which Schur polynomials
appear in $\phi^*(s_\l)$ is equivalent to asking which irreducible representations appear
in a certain tensor product of representations of the symmetric group.

While we are concerned with the action of $\phi^*$ on Schur polynomials acting on a fixed number
of variables, we will simplify our discussion by
working in the ring of symmetric functions.  Define a {\em symmetric function} to
be a set of symmetric polynomials $p(x_1, \ldots, x_l)$, one for each positive integer $l$,
such that
\begin{equation}
p(x_1, \ldots, x_l, 0, \ldots, 0) = p(x_1, \ldots, x_1).
\end{equation}
Recall that the Newton power sum symmetric functions are defined
as follows.  For a nonnegative integer $s$ (which we may also
think of as a partition of one part of size $s$), $p_s(X_1,
\ldots, X_k) = X_1^s + \cdots + X_k^s$.  For a partition $\l =
(\l_1, \ldots, \l_l)$ of length $l$, define
\begin{equation}
p_\l(X_1, \ldots, X_k) = \prod_{i=1}^l p_{\l_i}(X_1, \ldots, X_k).
\end{equation}
As we have seen, $\phi^*(p_s) = d_B p_s$, so
that $\phi^*(p_\l) = d_B^{l(\l)}p_\l$, where $l(\l)$ is the length
of the partition $\l$.

We use the following basic facts about the representation theory of the symmetric group
\cite{FH91, Fulton97}.  The irreducible representations of the symmetric group $S_n$
on $n$ letters can be put in one-to-one correspondence
with the partitions of $n$, in a standard way.  (And the partitions of $n$ also
correspond naturally to the conjugacy classes of $S_n$.) Furthermore, the Newton power sum symmetric
functions $p_\m$ and the Schur polynomials $s_\l$ are related as follows.
For any partition $\m$ of $n$, define
\begin{equation}
z(\m) = \prod_r r^{m_r}(m_r!),
\end{equation}
where $m_r$ is the number of times $r$ occurs in $\m$.
Now for any partition
$\m$ of $n$,
\begin{equation}
p_\m = \sum_\l \c_\m^\l s_\l;
\end{equation}
and for any partition $\l$ of $n$,
\begin{equation}
s_\l = \sum_\m \frac{1}{z(\m)}\c^\l_\m p_\m,
\end{equation}
where $\c_\m^\l$ is the character of the representation labelled by $\l$ evaluated on
a permutation in the conjugacy class labelled by $\m$.

Let us now return to the fact that $\phi^*(p_\l) = d_B^{l(\l)}
p_\l$.  This means that $\phi^*$ is a class function on $S_{d_B}$
(where $d_B = |\l|$), so we wish to find a representation $\r$ of
$S_{n}$ such that the character $\c^\r$ of $\r$ is equal to
$\phi^*$. Consider the representation $\r$ of $S_{n}$ on $B^{\ox
n}$ that acts by permuting the tensor factors: if
$\{e_i\}_{i=1}^{d_B}$ is an orthogonal basis for $B$, then for $w
\in S_{n}$,
\begin{equation}
\r(w)(e_{i_1} \ox \cdots \ox e_{i_{n}}) = e_{i_{w(1)}} \ox \cdots
\ox e_{i_{w(n)}}.
\end{equation}
We claim that the character $\c^\r = \phi^*$, or in other words,
for any $w \in S_{n}$, the character of $\r$ evaluated at $w$ is
$d_B^{l(w)}$, where $l(w)$ is the number of cycles in $w$.  To see
this, recall that by definition, $\c^\r(w) = \Tr(\r(w))$.  So
$\c^\r(w)$ is the number of elements of the basis $\{e_{i_1} \ox
\cdots \ox e_{i_{n}}\}$ fixed by the map $\r$; in other words,
\begin{equation}
\c^\r(w) = |\{(i_1, \ldots, i_{n}) = (i_{w(1)}, \ldots,
i_{w(n)})\}|.
\end{equation}
Now, if $(i_1, \ldots, i_{n})$ is fixed by $w$, then for any $r_1$
and $r_2$ in the same cycle of $w$, we must have $i_{r_1} =
i_{r_2}$.  Conversely, if $(i_1, \ldots, i_{d_B})$ satisfies the
property that $i_{r_1} = i_{r_2}$ for any $r_1$ and $r_2$ in the
same cycle of $w$, then  $(i_1, \ldots, i_{n})$ is fixed by $w$.
We conclude that the number of elements in the set $\{(i_1,
\ldots, i_{n}) = (i_{w(1)}, \ldots, i_{w(n)})\}$ is equal to the
number of ways to assign a basis element to each cycle of $w$,
which is $d_B^{l(w)}$.

Let $V_\l$ be the irreducible representation of $S_{n}$ labelled
by $\l$.  Let
\begin{equation}
V_\l \ox \r = \oplus_\p (V_\p)^{\ox m_\p}
\end{equation}
be a decomposition of $V_\l \ox \r$ into irreducible representations (each irrep $V_\p$ occurs
with multiplicity $m_\p$).  Then we have that
\begin{equation}
\c^\l(\m) \c^\r(\m) = \sum_\p m_\p \c^\p(m),
\end{equation}
a result we will use in the next calculation, the evaluation of
$\phi^*(s_\l)$:
\begin{eqnarray}
\phi^*(s_\l) & = & \sum_\m \frac{1}{z(\m)} \c^\l(\m)\phi^*(p_\m)\\
& = & \sum_\m \frac{1}{z(\m)} \c^\l(\m) d_B^{l(\m)} p_\m\\
& = & \sum_\m \frac{1}{z(\m)} \c^\l(\m) \c^\r(\m) p_\m\\
& = & \sum_\m \frac{1}{z(\m)} \sum_\p m_\p \c^\p(\m) (p_\m)\\
& = & \sum_\p \sum_\m \frac{1}{z(\m)} \c^\p(\m) (p_\m)\\
& = & \sum_\p m_\p s_\p.
\end{eqnarray}

So the Schur polynomials $s_\p$ appearing in $\phi^*(s_\l)$ are precisely those corresponding
to the representations $V_\p$ appearing in $V_\l \ox \r$.

\subsection{Sufficiency}
\label{sec:sufficiency}

We have described an approach using a variational principle to determine inequalities relating
a matrix $\r_{AB}$ to its partial trace $\r_A$, along with some observations for simplifying the
list of inequalities.
While our method has the advantage of relative straightforwardness and
simplicity, our techniques do not (to our knowledge) allow us to demonstrate that the inequalities
obtained are in fact sufficient: that is, if $\l$ and $\tilde \l$ satisfy the inequalities, then
there exists matrices $\r_{AB}$ and $\r_A = \Tr_B \r_{AB}$ such that $\l$ is the spectrum of
$\r_{AB}$ and $\tilde \l$ is the spectrum of $\r_A$.  It turns out that the inequalities obtained
from our variational principle approach are indeed sufficient.  This follows from recent work in
symplectic geometry \cite{BS98}, of which we became aware after deriving the inequalities through
our methods.  In this section, we will state the main result from \cite{BS98} and show that it yields
inequalities equivalent to the ones we have obtained.

We can express our problem in the language of symplectic geometry.
(See \cite{Knutson00, BS98, Silva01} for definitions and further
discussion.) Consider the Lie group $U(A \ox B)$ of unitary
matrices acting on the space $A \ox B$. For any vector $\l =
(\l_1, \ldots, \l_{d_A d_B})$ with terms arranged in nonincreasing
order, the set $\cO_\l^{AB}$ of Hermitian matrices on $A \ox B$
with spectrum $\l$ is a coadjoint orbit of $K = U(A \ox B)$.  Now
consider the action of the Lie group $\tilde K = U(A)$ of unitary
matrices on $A$, by conjugation on the symplectic manifold
$\cO_\l^{AB}$: for $U \in U(A)$,
\begin{equation}
U: \r_{AB} \mapsto (U \ox I_B) \r_{AB} (U^\dagger \ox I_B).
\end{equation}
It is not hard to verify that this is a Hamiltonian group action
whose moment map is $\Tr_B$, the partial trace with respect to
$B$.  So our problem, then, is to describe the image of the
symplectic manifold $\cO_\l^{AB}$ under the moment map $\Tr_B$.

This formulation is useful because considerable work has been done
in the study of the image of moment maps.  For instance, the
following result is due to Kirwan \cite{Kirwan,Knutson00}:

\begin{theorem}
Let $M$ be a compact connected Hamiltonian $K$-manifold, with moment map $\Phi$.
Then the intersection of the image of $\Phi$ with the positive Weyl chamber
${\mathfrak t}^*_+$ is a convex polytope.
\end{theorem}

In our case, the positive Weyl chamber of $U(A)$ consists of diagonal matrices
whose diagonal entries are in nonincreasing order (every matrix in the image of
$\Phi$ has the same spectrum as one such matrix).
Kirwan's theorem thus allows us to conclude that the set of all
ordered spectra of matrices obtainable by taking the partial
traces of matrices with a fixed spectrum must be a region bounded
by a finite set of inequalities.

Interestingly, Horn's problem can also be viewed in this framework.
Recall that Horn's problem asks for the possible spectra of $X + Y$, given the spectra of
$n \times n$ matrices $X$ and $Y$.  Suppose that $\l$ is the spectrum of $X$ and $\m$ is the
spectrum of $Y$.
Now we consider the action of the group $U(n)$ of $n \times n$ unitary matrices
 on the symplectic manifold
$\cO_\l \times \cO_\m$ by diagonal conjugation:
\begin{equation}
U: (X, Y) \mapsto (UXU^\dagger, UYU^\dagger).
\end{equation}
This is a Hamiltonian group action whose moment map takes two Hermitian matrices to their sum.
Thus, Horn's problem can be viewed as the problem of determining the image of this moment map.

The following theorem of Berenstein and Sjamaar \cite{BS98}
generalizes Klyachko's solution to Horn's problem. Before we state
it, some new notation is required. Let $K$ be a compact connected
Lie group, and let $\tilde K$ be a closed connected subgoup.  Let
$f$ be the inclusion map of $\tilde K$ into $K$, $f_*: \tilde \mfk
\rightarrow \mfk$ be the embedding of Lie algebras induced by $f$,
and $f^*: \mfkd \rar \tilde \mfkd$ be the dual projection.  Choose
maximal tori $T$ of $K$ and $\tilde T$ of $\tilde K$, and Weyl
chambers ${\mathfrak t}^*_+ \subset {\mathfrak t}^*$ and $\tilde
{\mathfrak t}^*_+ \subset \tilde {\mathfrak t}^*$, where
${\mathfrak t}$ and $\tilde {\mathfrak t}$ are the Lie algebras of
$T$ and $\tilde T$, respectively. For $\a \in {\mathfrak t}^*_+$,
let $\Delta(\cO_\a) = f^*(\cO_\a) \cap \tilde {\mathfrak t}^*_+$.
Let $\cC$ be the cone spanned by the simple roots of $\mathfrak
t^*$.  Let $W$ and $\tilde W$ be the Weyl groups of $K$ and
$\tilde K$ respectively.  Let $\phi$ be the embedding of the flag
variety $\tilde K / \tilde T$ into the flag variety $K/T$ which is
induced by the map $f$.  We now state the main result from
\cite{BS98}:
\begin{theorem}
\label{thm:bs}
Let $(\tilde \a, \a) \in \tilde {\mathfrak t}^*_+ \times {\mathfrak t}^*_+$.
Then $\tilde \a \in \Delta(\cO_\a)$ if and only if
\begin{equation}
\label{eqn:cone}
\tilde w^{-1} \tilde \a \in f^*(w^{-1} \a - v\cC)
\end{equation}
for all triples $(\tilde w, w, v) \in \tilde W \ox W \ox W_{\mbox{rel}}$ such that
$\phi^*(v\s_{wv})(\tilde c_{\tilde w}) \not = 0$.
\end{theorem}
(Here $W_{\mbox{rel}}$ is the {\em relative Weyl set}, defined in
\cite{BS98}.  We shall not be concerned with the details of its
description; it is equal to $\{1\}$ for our case.) For any $w \in
W$, $f^*(w^{-1} \l - \cC)$ is a polyhedral cone in $\tilde
{\mathfrak t}^*$, so Equation~(\ref{eqn:cone}) represents a finite
number of inequalities.  The theorem gives us inequalities
whenever the condition $\phi^*(v\s_{wv})(\tilde c_{\tilde w}) \not
= 0$ is satisfied, where $\s_{wv}$ is the element of the
cohomology of the flag variety labelled by Weyl group element
$wv$, and $\tilde c_{\tilde w}$ is the element of the homology of
the flag variety labelled by $\tilde w$. This is equivalent to the
condition that $\tilde \s_{\tilde w}$ appears in $\phi^*(\s_w)$,
remembering that $v = 1$ for us.

We review some facts about the cohomology of flag varieties of a complex
vector space $V$ \cite{Fulton97}.  Fix a flag $F_\bullet$ of $V$.  The cohomology
classes $\s_{w}$, known as {\em Schubert classes},
 are indexed by elements of $S_n$, where $n = \dim V$.  For
$w \in S_n$, $\s_w$ corresponds to the class of the {\em Schubert variety} $X_w$,
which is the closure of the {\em Schubert cell}
\begin{equation}
\O_w = \{ E_\bullet \in \Fl (V)| \dim(E_p \cap F_q) = \#\{i \leq
p: w(i) \leq q\} \mbox{ for } 1 \leq p, q \leq m\}.
\end{equation}

Let us specialize to the case of our problem of finding the
spectrum of a partial trace. For this case $f^*(\cC) = \tilde
\cC$. If $\tilde \s_{\tilde w}$ appears in $\phi^*(\s_w)$,
Equation~(\ref{eqn:cone}) tells us that
\begin{equation}
f^*(w^{-1} \a) - \tilde w^{-1} \tilde \a \in  \tilde \cC
\end{equation}
for elements of the dual space $\a \in {\mathfrak t}^*_+, \tilde \a \in \tilde
{\mathfrak t}^*_+$.  These functionals $\a$, $\tilde \a$ act on the spectra
$\l$, $\tilde \l$; we have
\begin{equation}
(w^{-1}\a) (\l) = \a(w^{-1}(\l)) = \a(\l_{w(1)}, \l_{w(2)}, \ldots, \l_{w(n)}).
\end{equation}
Identifying $\mathfrak t$ and $\tilde {\mathfrak t}$ with their dual spaces,
we have the conditions that
\begin{equation}
f^*(\l_{w(1)}, \l_{w(2)}, \ldots, \l_{w(d_A d_B)}) -  (\l_{\tilde w(1)}, \l_{\tilde w(2)},
\ldots, \l_{\tilde w(d_A)}) \in \tilde \cC
\end{equation}
whenever $\tilde \s_{\tilde w}$ appears in $\phi^*(\s_w)$.  But
the root cone $\cC$ is generated by the simple
roots $\l_1 - \l_2, \l_2 - \l_3, \ldots, \l_{d_A - 1} - \l_{d_A}$ where $\l_i \geq \l_{i+1}$;
in order words, $\cC$ is generated by the set of $\m$ such that
\begin{equation}
\sum_{i=1}^k \m_i \geq 0, \mbox{ for } k < d_A,
\end{equation}
and
\begin{equation}
\sum_{i=1}^{d_A} \m_i = 0.
\end{equation}
So our conditions are that
\begin{equation}
(0, 0, \ldots, 0) \prec
f^*(\l_{w(1)}, \l_{w(2)}, \ldots, \l_{w(d_A d_B)}) -  (\l_{\tilde w(1)}, \l_{\tilde w(2)},
\ldots, \l_{\tilde w(d_A)}),
\end{equation}
or
\begin{eqnarray*}
\lefteqn{(0, 0, \ldots, 0) \prec } \\
& & (\l_{w(1)} + \ldots + \l_{w(d_B)}, \l_{w(d_B + 1)} + \ldots + \l_{w(2 d_B)}, \ldots,
\l_{(w((d_A -1)d_B + 1)} + \ldots + \l_{w(d_A d_B)}) \\
& & -  (\l_{\tilde w(1)}, \l_{\tilde w(2)},
 \ldots, \l_{\tilde w(d_A)}).
\end{eqnarray*}
This is turn yields $(d_A - 1)$ inequalities:
\begin{eqnarray}
\sum_{i=1}^{d_B} \l_{w(i)}  & \leq & \tilde \l_{\tilde w(1)}, \\
\sum_{i=1}^{2d_B}\l_{w(i)} & \leq & \tilde \l_{\tilde w(1)} + \l_{\tilde w(2)}, \\
& \vdots & \\
\sum_{i=1}^{(d_A -1) d_B}&  \leq & \sum_{i=1}^{d_A -1}   \tilde \l_{\tilde w(i)}.
\end{eqnarray}
These inequalities arise from intersections of Schubert cells of
the flag varieties but any such inequality can be obtained as a
consequence of an intersection of Grassmannian Schubert varieties.
Choose a flag variety $F_\bullet$ of $A \ox B$ corresponding to the eigenspaces of
$\r_{AB}$ arranged in nonincreasing order of eigenvalues, and a flag variety
$\tilde F_\bullet$ of $A$ corresponding to the eigenspaces of
$\r_A$ arranged in nonincreasing order of eigenvalues.  Now define $\p_w$ to be the binary
string of length $d_A d_B$ such that
\begin{eqnarray*}
\p_w(i) = 1 & \mbox{ if } w^{-1}(i) \leq kd_B, \\
\p_w(i) = 0 & \mbox{ otherwise}.
\end{eqnarray*}
Similarly, define $\tilde \p_{\tilde w}$ to be the binary string of length $d_A$
which takes on the value $1$ only at those positions $i$ such that $\tilde w^{-1}(i) \leq k$.

Now consider any inequality of the form
\begin{equation}
\sum_{i=1}^{kd_B}\l_{w(i)} \leq \sum_{i=1}^k   \tilde \l_{\tilde w(i)},
\end{equation}
for some permuations $w$ and $\tilde w$, arising from the intersection of
$\O_w(F_\bullet)$ and $\phi(\O_{\tilde w}(\tilde F_\bullet))$.
Suppose $E_\bullet \in \O_w(F_\bullet)$ and
$\tilde E_\bullet \in \O_{\tilde w}(\tilde F_\bullet)$, such
that $\phi(\tilde E_\bullet) = E_\bullet$.
Therefore, the subspaces $E_{nk}$ and
$\tilde E_k$ satisfy
$\phi(\tilde E_k) = E_{nk}$.  Note that
$E_{nk} \in \O_{\p_w}(F_\bullet)$, and
$\tilde E_k \in \O_{\p_{\tilde w}}(\tilde F_\bullet)$,
where $\O_{\p_w}(F_\bullet)$ and  $\O_{\p_{\tilde w}}(\tilde F_\bullet)$
are Grassmannian Schubert cells.  Therefore, we have a nonempty intersection
 $\O_{\p_w}(F_\bullet) \cap \phi(\O_{\p_{\tilde w}}(\tilde F_\bullet)) \not = \emptyset$,
which by Theorem~\ref{thm:varprin} yields the same inequality
\begin{equation}
\sum_{i=1}^{kd_B}\l_{w(i)} \leq \sum_{i=1}^k   \tilde \l_{\tilde w(i)}.
\end{equation}
Thus, considering only Grassmannian intersections is enough to
derive any inequality of Theorem~\ref{thm:bs} applied to our
problem. So the inequalities derived by the approach we have
described are indeed sufficient.

\subsection{The large message limit}

Having determined how to find the inequalities relating $\r_{AB}$
and $\r_A$, we can seek methods of simplifying the list of
inequalities.  It turns out that the inequalities governing the
relationship between the spectra of $\r_{AB}$ and of $\r_A$ are
particularly simple when $d_B$ is large compared to $d_A$.   In
this section we will show that if $d_B \geq \frac{1}{2} d_A^2$,
then the basic inequalities are sufficient. (All other
inequalities are redundant.) Physically, thinking in terms of a
quantum communication protocol where Alice sends $\log_2 d_B$
qubits to Bob, such a result is plausible because a large amount
of communication gives Alice a great deal of freedom in
manipulating her portion of the system, so we should not expect
there to be much restriction in the states she might end up with.

Suppose that $d_B \geq \frac{1}{2}d_A^2$, and consider an
arbitrary inequality resulting from the nonzero cup product
$\tilde \s_\n \cup \phi^*(\s_\p) \not = 0$. (As discussed in
Section~\ref{sec:together}, we may assume that $\tilde \s_\n$ is a
summand in the expansion of $\phi^*(\s_\p) \not = 0$ as a sum of
Schubert classes.) Such an inequality is of the form
\begin{equation}
\label{eqn:satineq}
\sum_{i \in I} \tilde \lambda_i \leq \sum_{j \in J} \lambda_j
\end{equation}
where if $|I| = k$, then $|J| = d_B k$.  As in
Section~\ref{sec:obs}, we may assume that $k \leq \frac{d_A}{2}$.
Consider the partitions $\p$ and $\n$ in the equation $\tilde
\s_\n \cup \phi^*(\s_\p) \not = 0$ to be binary strings. Let $u$
be the $(0,1)$ vector of length $d_A d_B$, whose $i$th component
is equal to $1$ if and only if $\p(i) = 1$.  Similarly, let
$\tilde u$ be the $(0,1)$ vector of length $d_A$, whose $i$th
component is equal to $1$ if and only if $\n(i) = 1$.  Then
Inequality~(\ref{eqn:satineq}) can be rewritten as
\begin{equation}
\label{eqn:satineq2}
\tilde \l \cdot \tilde u \leq \l \cdot u.
\end{equation}
We now prove some easy facts about this situation, ending with our
desired result.
\begin{observation}
\label{obs:boxes}
The Young diagram corresponding to $\p$ can't have more than $({d_A \over 2})^2$ boxes.
\end{observation}
This follows because the Young diagram corresponding to $\n$ must fit in a
$k \times ({d_A -k})$ rectangle, and so cannot have more than $({d_A \over 2})^2$ boxes;
and $\p$ must have the same number of boxes in its Young diagram as $\n$. \hfill $\Box$

\begin{observation}
\label{obs:dotprod}
If $u \prec u'$, then $\l \cdot u \leq \l \cdot u'$.
\end{observation}
This follows easily from the fact that $\l$ has its terms arranged in nonincreasing order.
\hfill $\Box$

\begin{observation}
\label{claim:box1} If $j > d_B k + ({d_A \over 2})^2$, then $j
\not \in J$ in Inequality~(\ref{eqn:satineq}) (in other words,
$\l_j$ is not one of the terms in the right hand sum).
\end{observation}
{\bf Proof \/}
If $j \in J$, then the
Young diagram corresponding to $\p$ would have more than $({d_A \over 2})^2$ boxes in its $j$th
row. \hfill $\Box$

\begin{observation}
\label{claim:box2}
 The first zero of $u$ can't appear before the $(d_B k - \lfloor ({d_A \over 2})^2 \rfloor)$th component.
In other words, if $j \leq d_B k - ({d_A \over 2})^2$, then $j \in
J$ in Inequality~(\ref{eqn:satineq}).
\end{observation}
{\bf Proof \/} Otherwise, the Young diagram corresponding to $\p$ would have more than $({d_A \over 2})^2$ rows.
\hfill $\Box$

\begin{lemma}
\label{lemma:maj}
\begin{equation}
(\underbrace{1, \ldots, 1}_{d_B k- \lfloor ({d_A \over 2})^2 \rfloor},
\underbrace{0, \ldots, 0}_{\lfloor ({d_A \over  2})^2\rfloor},
\underbrace{1, \ldots, 1}_{\lfloor ({d_A \over 2})^2 \rfloor},
0, \ldots, 0) \prec u.
\end{equation}
Consequently, since $d_B \geq \frac{d_A^2}{2}$,
\begin{equation}
(\underbrace{1, \ldots, 1}_{d_B k- \lfloor{d_B \over 2} \rfloor},
\underbrace{0, \ldots, 0}_{\lfloor {d_B \over  2}\rfloor},
\underbrace{1, \ldots, 1}_{\lfloor {d_B \over 2} \rfloor},
0, \ldots, 0)  \prec u.
\end{equation}
\end{lemma}
{\bf Proof \/} This follows from Observations~\ref{claim:box1}
and~\ref{claim:box2}. \hfill $\Box$

\begin{theorem}
If $d_B \geq \frac{1}{2}d_A^2$, then
Inequality~(\ref{eqn:satineq2}) is redundant.  In other words, the
basic inequalities are sufficient to characterize the relationship
between the spectrum of $\r_{AB}$ and the spectrum of $\r_A$.
\end{theorem}
{\bf Proof \/}
It is sufficient to assume that
$\tilde u \prec (\underbrace{1, \ldots, 1}_{k-1}, 0, 1,
0, \ldots, 0)$ (the only possible $\tilde u$ that does not
satisfy this condition is $\tilde u = (\underbrace{1, \ldots, 1}_k, 0, \ldots, 0)$,
which gives rise to the basic inequalities).
Then we have
\begin{eqnarray*}
\tilde \lambda \cdot \tilde u & \leq & \lambda \cdot (\underbrace{1, \ldots, 1}_{k-1}, 0, 1,
0, \ldots, 0)\\
& = & \sum_{i = 1}^{k-1} \tilde \lambda_i
+ \tilde \lambda_{k+1} \\
& \leq & {1 \over 2} \left[\sum_{i = 1}^{k-1} \tilde \lambda_i
+ \tilde \lambda_k + \tilde \lambda_{k+1} + \sum_{i =1}^{k-1} \tilde \lambda_i \right] \\
& = & (\underbrace{1, \ldots, 1}_{k-1}, {1 \over 2}, {1 \over 2},
0, \ldots, 0) \cdot \tilde \lambda \\
& = & {1 \over 2}(\underbrace{1, \ldots, 1}_{k-1}, 0, \ldots, 0) ^. \tilde \lambda +
 {1 \over 2}(\underbrace{1, \ldots, 1}_{k+1}, 0, \ldots, 0) ^. \tilde \lambda \\
& \leq & {1 \over 2}(\underbrace{1, \ldots, 1}_{d_B(k-1)}, 0, \ldots, 0) ^. \lambda  +
  {1 \over 2}(\underbrace{1, \ldots, 1}_{d_B(k+1)}, 0, \ldots, 0) ^. \lambda \quad \mbox{ by the basic inequalities}\\
& = & (\underbrace{1, \ldots, 1}_{d_B (k-1)},
\underbrace{{1 \over 2}, \ldots, {1 \over 2}}_{2 d_B}, 0, \ldots, 0) ^. \lambda \\
& \leq & (\underbrace{1, \ldots, 1}_{d_B k - \lfloor {d_B \over 2} \rfloor},
\underbrace{0, \ldots, 0}_{\lfloor {d_B \over 2} \rfloor},
\underbrace{1, \ldots, 1}_{\lfloor {d_B \over 2} \rfloor}, 0, \ldots, 0) ^. \lambda.
\end{eqnarray*}
But the right hand side of Inequality~(\ref{eqn:satineq2}) must be
greater than equal to
\begin{equation*}
(\underbrace{1, \ldots, 1}_{d_B k - \lfloor {d_B \over 2} \rfloor},
\underbrace{0, \ldots, 0}_{\lfloor {d_B \over 2} \rfloor},
\underbrace{1, \ldots, 1}_{\lfloor {d_B \over 2} \rfloor}, 0, \ldots, 0) ^. \lambda,
\end{equation*}
by Lemma~\ref{lemma:maj} and Observation~\ref{obs:dotprod}.  Thus,
we have shown that Inequality~(\ref{eqn:satineq2}) must hold,
assuming only the basic inequalities; so this inequality must be
redundant, for an arbitrary inequality arising from
 $\tilde \s_\n \cup \phi^*(\s_\p) \not = 0$.
\hfill $\Box$

We conjecture a stronger result, which we have verified for $d_A =
2, 3$, and $4$. (The cases $d_A = 2$ and $d_A = 3$ have been shown
explicitly in this paper.)

\begin{conjecture}
If $d_B \geq d_A$, then the basic inequalities are sufficient to
characterize the relationship between the spectrum of $\r_{AB}$
and the spectrum of $\r_A$.
\end{conjecture}

\section{Conclusion}

We have seen that the question of whether a particular quantum
state transformation can be accomplished with a given finite
amount of communication, classical or quantum, can be answered by
testing a set of inequalities determined by a cohomological
condition. In the classical communication case, the question
essentially reduces to Horn's Problem. We found, however, that
there is a simplification in the sense that all matrices can be
assumed to be isospectral or, in communication language, all
messages equiprobable. The case of state transformations using
quantum communication and only unitary local operations was found
to be amenable to a similar analysis but the cohomological
condition was different. Nonetheless, in the limit that the amount
of communication is large relative to the size of the state kept
behind, a significant simplification occurred, reducing the
complicated set of inequalities to a type of majorization. The
techniques presented here, in particular the theorem of Berenstein
and Sjamaar, are applicable to wide range of problems in linear
algebra. It is our hope that they will find further applications
in quantum information theory.

\section*{Acknowledgments}

The authors would like to thank Sergey Bravyi, Marco Gualtieri,
Alexander Klyachko, Allen Knutson, Michael Nielsen, John Preskill,
Eric Rains and Terrence Tao for various forms of help. Eric Rains,
in particular, illuminated us on the connection to representation
theory in Section \ref{subsec:repnTheory} and Allen Knutson spent
many patient hours teaching us about symplectic geometry, among
other things.

This paper is based on SD's Caltech doctoral
thesis~\cite{Daftuar03}. SD and PH are supported by the NSF
through grant EIA-0086038. PH is also grateful for funding from
the Sherman Fairchild Foundation and the Canadian Institute for
Advanced Research.

\bibliographystyle{plain}
\bibliography{schubert}

\end{document}

%% file: loqc.eepic
\setlength{\unitlength}{0.00057in}
\begingroup\makeatletter\ifx\SetFigFont\undefined%
\gdef\SetFigFont#1#2#3#4#5{%
  \reset@font\fontsize{#1}{#2pt}%
  \fontfamily{#3}\fontseries{#4}\fontshape{#5}%
  \selectfont}%
\fi\endgroup%
{\renewcommand{\dashlinestretch}{30}
\begin{picture}(6474,7350)(0,-10)
\put(312,6975){\ellipse{424}{424}}
\path(312,6750)(312,6225)(312,6300)
\path(312,6525)(12,6375)
\path(312,6525)(612,6375)
\path(312,6225)(87,5925)
\path(87,5925)(462,5925)
\path(312,6225)(537,5925)
\path(462,5925)(537,5925)
\path(237,5925)(237,5625)
\path(387,5925)(387,5625)
\put(6162,6975){\ellipse{424}{424}}
\path(6162,6750)(6162,6000)
\path(6162,6000)(5862,5625)
\path(6162,6000)(6387,5625)
\path(6162,6450)(5862,6225)
\path(6162,6450)(6462,6225)
\thicklines
\path(1362,6825)(2037,6825)(2037,6075)
    (1362,6075)(1362,6825)
\path(1362,5250)(2037,5250)(2037,4500)
    (1362,4500)(1362,5250)
\path(4512,6000)(5187,6000)(5187,5250)
    (4512,5250)(4512,6000)
\path(4512,4275)(5187,4275)(5187,3525)
    (4512,3525)(4512,4275)
\thinlines
\path(2262,6375)(4287,5700)
\blacken\path(4163.671,5709.487)(4287.000,5700.000)(4182.645,5766.408)(4163.671,5709.487)
\path(4287,5475)(2262,4875)
\blacken\path(2368.533,4937.855)(2262.000,4875.000)(2385.578,4880.327)(2368.533,4937.855)
\path(2254,4703)(4279,4028)
\blacken\path(4155.671,4037.487)(4279.000,4028.000)(4174.645,4094.408)(4155.671,4037.487)
\thicklines
\path(1362,2100)(2037,2100)(2037,1350)
    (1362,1350)(1362,2100)
\thinlines
\path(2262,1650)(4287,975)
\blacken\path(4163.671,984.487)(4287.000,975.000)(4182.645,1041.408)(4163.671,984.487)
\thicklines
\path(4512,1275)(5187,1275)(5187,525)
    (4512,525)(4512,1275)
\thinlines
\path(4354,2426)(2329,1826)
\blacken\path(2435.533,1888.855)(2329.000,1826.000)(2452.578,1831.327)(2435.533,1888.855)
\put(3000,7300){$|\varphi_{AB}\rangle$}
\put(1350,7000){$\overbrace{\quad\quad\quad\quad\quad\quad\quad\quad\quad\quad\quad\quad\quad\quad}$}
\put(1587,4800){$U_3$}
\put(4737,3825){$U_4$}
\put(4737,825){$U_k$}
\put(4737,5550){$U_2$}
\put(1630,2975){$\vdots$}
\put(3250,2975){$\vdots$}
\put(4800,2975){$\vdots$}
\put(3162,6225){$q_1$}
\put(3162,5325){$q_2$}
\put(3162,4500){$q_3$}
\put(3020,2275){$q_{k-1}$}
\put(3162,1425){$q_k$}
\put(1420,1650){$U_{k-1}$}
\put(1587,6375){$U_1$}
\put(1350,450){$\underbrace{\quad\quad\quad\quad\quad\quad\quad\quad\quad\quad\quad\quad\quad\quad\;}$}
\put(3000,0){$|\psi_{AB}\rangle$}
\end{picture}
}